\documentclass[preprint,aps,showpacs,pre,amsmath,amssymb]{revtex4-1}

\usepackage{amssymb}
\usepackage{graphicx}
\usepackage{dsfont}
\usepackage{color,soul}
\usepackage{xcolor}
\usepackage{appendix}

\begin{document}
	
\title{Recursive Green's functions optimized for \\ atomistic modeling of large superlattice-based devices}
	
	\author{V. Hung Nguyen and J.-C. Charlier}
	\affiliation{Institute of Condensed Matter and Nanosciences, Universit\'{e} catholique de Louvain (UCLouvain), Chemin des \'etoiles 8, B-1348 Louvain-la-Neuve, Belgium}

\begin{abstract}
The Green's function method is recognized to be a very powerful tool for modeling quantum transport in nanoscale electronic devices. As atomistic calculations are generally expensive, numerical methods and related algorithms have been developed accordingly to optimize their computation cost. In particular, recursive techniques have been efficiently applied within the Green's function calculation approach. Recently, with the discovery of Moiré materials, several attractive superlattices have been explored using these recursive Green's function techniques. However, numerical difficulty issues were reported as most of these superlattices have relatively large supercells, and consequently a huge number of atoms to be considered. In this article, improvements to solve these issues are proposed in order to keep optimizing the recursive Green's function calculations. These improvements make the electronic structure calculations feasible and efficient in modeling large superlattice-based devices. As an illustrative example, twisted bilayer graphene superlattices are computed and presented to demonstrate the efficiency of the method.
\end{abstract}

\maketitle
	
\section{Introduction}\label{sec1}

For several decades, numerical modeling based on atomistic simulation has played a prominent role in designing and understanding the electronic properties of nanomaterials and their corresponding devices. Different simulation toolboxes have been developed particularly, based on classical (e.g., drift-diffusion \cite{Granzner2006}), semi-classical (Kubo-Greenwood \cite{Fan2014,Fan2021}, Monte-Carlo \cite{Jacoboni1983}, deterministic solution of Boltzmann transport equation \cite{Hong2009}), and quantum transport methods (e.g., non-equilibrium Green’s functions (NEGF) \cite{Datta1995,Mingo2007,Pourfath2014}). Among these methods, the NEGF approach presents important advantages in modeling of nanoscale devices, in order to address the importance of quantum confinement effects induced by the device's finite size. 

NEGF has been primarily developed for large bias (out-of-equilibrium) calculations. It then has been extended to compute also the low-bias transport (e.g., conductance \cite{Lewenkopf2013}, carrier mobility \cite{Niquet2014}, and so on). Very importantly, this technique can efficiently address the interplay between quantum confinement effects and scatterings (e.g., surface disorders, defects, impurities, phonons, etc.). Although atomistic simulations are generally expensive, NEGF calculations have benefited a lot from the progresses in numerical methods and algorithms \cite{Lake1997,Svizhenko2002,Li2007,Kazymyrenko2008,Anantram2008,Stephen2011,Lewenkopf2013,Do2014,Thorgilsson2014,Zhang2019} and from the development of high performance computing systems. The NEGF approach, therefore, is able to model very realistic systems and has important contribution to the research of nanomaterials and devices \cite{Mathieu2006,Bescond2007,Mathieu2009,Mathieu2010,Fiori2013,Alarcon2013,Nguyen2013,Bescond2014,Nguyen2014b,Bescond2017,Zhang2017,Choukroun2019}. In particular, it can be used to solve different electronic models such as effective mass, \textbf{\textit{k.p}} and atomistic tight-binding Hamiltonians \cite{Mathieu2006,Bescond2007,Mathieu2009,Mathieu2010,Fiori2013,Alarcon2013,Nguyen2013,Bescond2014,Nguyen2014b,Bescond2017,Zhang2017,Choukroun2019,Bescond2004,Antonio2007,Mathieu2009,Groth2014,Nguyen2018,Brun2019,Nguyen2020}. NEGF calculations have even been implemented in density-functional-theory codes \cite{Ozaki2010,Papior2017} to compute \textbf{\textit{ab initio}} quantum transport. Moreover, they have been also extended to solve the equation of atomic vibration and therefore are able to compute quantum thermal transport \cite{Mingo2007,Zhang2007,Jinghua2009,Mazzamuto2011,Mazzamuto2012,Nguyen2014,Wang2014}.

The recursive method is  one of the numerical techniques that can improve significantly the efficiency of NEGF codes \cite{Lake1997,Svizhenko2002,Anantram2008,Stephen2011,Lewenkopf2013,Do2014,Thorgilsson2014,Zhang2019}. Its specificity consists in the ability to compute recursively the small-size blocks (instead of solving the full matrix) of Green's functions. Therefore, recursive Green's function codes can have a low computation cost, especially as they can be easily parallelized \cite{Drouvelis2006}. Such advantage is actually obtained when the long-range electronic couplings are negligible and therefore the system Hamiltonian can be numerically represented by a tridiagonal-block matrix form, which happens in most electronic devices. 

With the exploration of recent new materials (most remarkably, 2D layered materials) \cite{Avouris2017,Ferrari2015}, NEGF calculations however face some new computation issues. In particular, 2D materials offer a platform allowing the creation of heterostructures with a huge set of interesting and tunable properties.  
The most notable 2D (or quasi-2D) systems are van der Waals heterostructures \cite{Geim2013,Novoselov2016} and more recently moiré superlattices \cite{Feng2021,Wang2019,Andrei2020}, where 2D monolayers are hold together by van der Waals long-range weak forces. Moiré superlattices can be created by stacking monolayers of different materials presenting a non-negligible lattice mismatch and/or by relatively rotating certain layers in a multilayer stack (i.e., creating a twisted system). The periodic length of the corresponding superlattices is tunable as it is a function of the rotation (i.e., twist) angle and can reach tens to hundreds of nanometers (e.g., see in Refs. \cite{Yoo2019,Sunku2018}). When reaching such a large scale, electronic structure calculations of moiré supperlattices are extremely heavy and even unfeasible, although the recursive techniques reported in the literature are applied. In such a context, the current article will present some novel aspects for recursive calculations, aiming to solve the above-mentioned issues. The article is organized as follows. First, we will review briefly the atomistic NEGF simulations and the recursive techniques reported in the literature. The derivation of the recursive equations is explained using simple matrix operations. We will then present new calculations to solve the main obstacles (particularly, to optimize the calculations of contact self-energies) when modeling large superlattices. In the following section, novel recursive calculations will be proposed to solve the Green's function of periodic systems, which can be considered as an alternative approach to model their electronic structure. Finally, we will illustrate and discuss the efficiency of these newly developed methods using twisted bilayer graphene superlattices as an illustrative example.

\section{Atomistic modeling using Green's function method}\label{sec2}

\begin{figure}[ht]%
\centering
\includegraphics[width=0.9\textwidth]{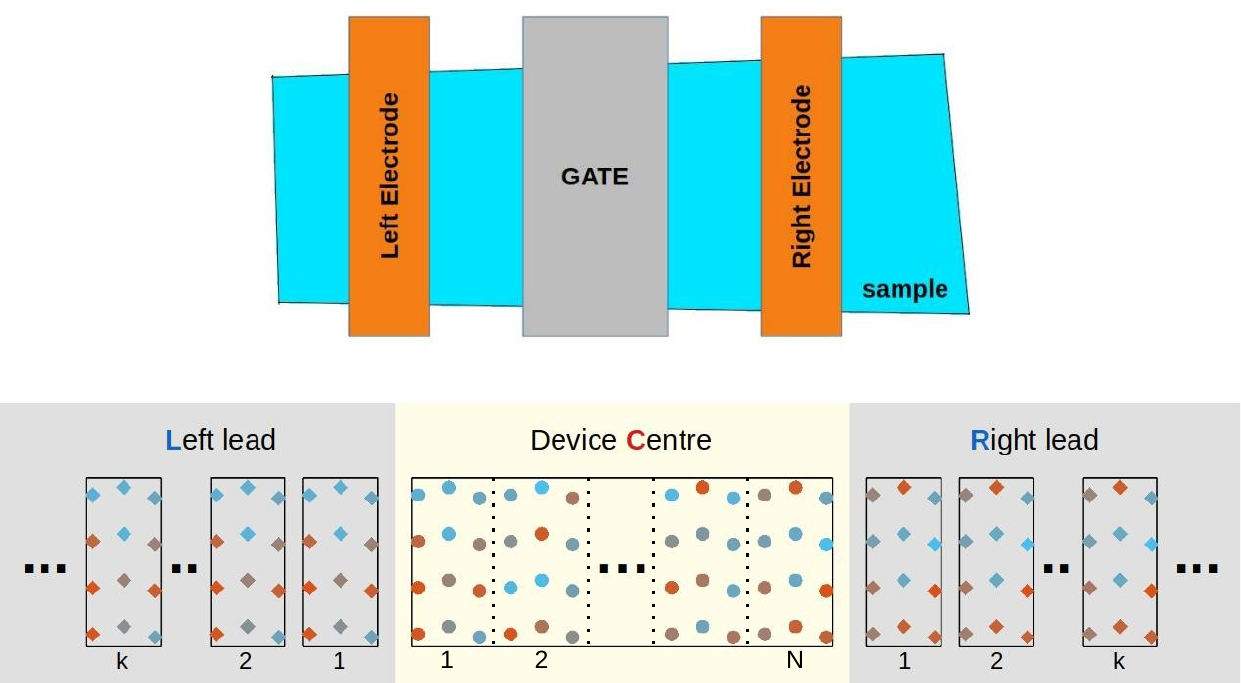}
\caption{Schematic of experimental setups for measuring electronic transport in a sample (top) and corresponding atomic structures for atomistic simulations (bottom).}\label{figS01}
\end{figure}

In practice, most electronic devices consist of a long conducting channel and external (typically, source and drain) electrodes patterned on their two ends as schematized on the top panel of Fig.\ref{figS01}. Gate electrodes can be deposited on the device centre to control the electronic transport through the system. To simulate this two-terminal device, the typical atomic model shown on the bottom panel of Fig.\ref{figS01} has been often considered in atomistic simulations. In particular, the devices can generally be divided into three parts: left lead, device centre, and right lead. The device centre is simply the centre part of realistic devices, where scatterings and controlability (e.g., by gates) take place and play central roles on the device performance. Two leads are used to model the extended (generally, large) parts at the two ends of the device, where, as described, source and drain electrodes can be patterned. In principle, to compute accurately the electronic properties of the leads and their couplings to the device, all possible scatterings (e.g., defects, impurities, phonons, etc.) as well as effects related to the contacting interface between the conducting sample and electrodes have to be taken into account. However, because of their expensive calculations, these effects have often been neglected or computed using the effective electronic models of the leads (e.g. see Refs. \cite{Pei2011,Salvador2012,Do2012,Houssa2019}). Therefore, in atomistic simulations, both two leads have generally been modeled using periodic and semi-infinite materials and can be feasibly computed using the developed numerical methods \cite{Lewenkopf2013,Do2014}. In summary, to fully explore the device performance, the electronic properties of the channel material, device-to-contact couplings, scattering effects in the device centre region as well as the size-dependent effects, etc. can be computed.

As early mentioned, the NEGF method has been demonstrated to be a powerful tool to investigate the electronic transport and quantum effects in nanoscale devices. Its efficiency is significantly enhanced when the recursive technique is applied. 
Indeed, this method is actually based on computing the retarded ($G^r$), advanced ($G^a$), lesser ($G^<$) and greater ($G^<$) Green's functions. Since $G^a$ and $G^\lessgtr$ can be computed using their relationship with the retarded one (named $G$ for short, hereafter), only the recursive calculations for $G$ will be discussed in this paper. Once these Green's functions are obtained, the electronic quantities as local density of states (LDOS), local density of charges (LDOC), transmission function, and so on, can be computed as explained in the following sections.

The retarded Green's function ($G_{sys}$) of the system at a given energy $E$ is the solution of the present equation \cite{Datta1995,Mingo2007,Pourfath2014}:
\begin{equation}
[E + i0^+ - H_{sys}] G_{sys} = \mathbb{I}.
\end{equation}
"$0^+$" is a positive inﬁnitesimal energy added to ensure the convergence of Green’s function calculations. The Hamiltonian $H_{sys}$ in the matrix form is written as
\begin{eqnarray}
H_{sys} = \begin{bmatrix}
H_L & H_{LC} & 0\\ 
H_{CL} & H_C & H_{CR}\\ 
0 & H_{RC} & H_R
\end{bmatrix}, 
\mathrm{\,and \, accordingly,} \hspace{0.4cm} G_{sys} = \begin{bmatrix}
G_L & G_{LC} & G_{LR}\\ 
G_{CL} & G_C & G_{CR}\\ 
G_{RL} & G_{RC} & G_R
\end{bmatrix}.
\end{eqnarray}
Here, $H_{L}$, $H_{R}$ and $H_C$ are the Hamiltonian matrices of the left lead, the right lead and the device centre, respectively, and $H_{L(R)C}$ models their couplings.
Using the formulas (\ref{eq:A1},\ref{eq:A2}) in Appendix A, the device Green's function $G_C$ can be computed by solving the equation
\begin{eqnarray}
A_C G_C &=& \mathbb{I}, \hspace{0.2cm} \mathrm{with} \hspace{0.2cm}
 A_C = E + i0^+ - H_C - \Sigma_L - \Sigma_R \label{eq:T2} \\
\mathrm{and} \hspace{0.2cm} \Sigma_{L(R)} &=& H_{CL(R)}G^0_{L(R)} H_{L(R)C}, \hspace{0.5cm} G^0_{L(R)} = (E+i0^+ - H_{L(R)})^{-1}. \nonumber
\end{eqnarray}
$\Sigma_{L (R)}$ are left (right) device-to-lead self-energies (i.e., contact self-energies, hereafter) induced by the device-lead couplings and $G^0_{L (R)}$ are accordingly the Green's function of isolated left (right) leads. As the device centre interacts only with atoms near the right (left) end of the left (right) lead, the self energies $\Sigma_{L,R}$ can be determined using the Green's function associated to these surface atoms in the leads. 

In general, the system can be partitioned into small slices along the transport direction as as illustrated in the bottom panel of Fig.\ref{figS01}, satisfying that there is only electronic coupling between the first nearest-neighboring slices. In such condition, the Hamiltonian matrix can be written in the tridiagonal-block form, leading the equation of $G^0_{L,R}$ to be
\begin{eqnarray}
&& \begin{bmatrix}
A^{L,R}_{11} & A^{L,R}_{12} & 0 & \cdots & \cdots & \cdots & \\ 
A^{L,R}_{21} & A^{L,R}_{22} & A^{L,R}_{23} & 0 & \cdots & \cdots & \\ 
0 & A^{L,R}_{32} & \cdots & \cdots & \cdots & \cdots & \\ 
\cdots & 0 & \cdots & \cdots & \cdots & \cdots & \\ 
\cdots & \cdots & \cdots & \cdots & \cdots & \cdots & \\ 
 &  &  &  &  &  & A^{L,R}_{N_{L,R},N_{L,R}}
\end{bmatrix} G^0_{L,R} = \mathbb{I}, \label{eq:T3} \\ \nonumber \\
\mathrm{with} \hspace{0.1cm} && A^{L,R} = E+i0^+ - H_{L,R}. \nonumber
\end{eqnarray}
When the leads are assumed to be periodic and the mentioned slices are identical to their unit cells, $A^{L,R}_{kk} \equiv A^{L,R}_{11}$,  $A^{L,R}_{k,k+1} \equiv A^{L,R}_{12}$, $A^{L,R}_{k,k-1} \equiv A^{L,R}_{21}$ is obtained \hspace{0,1cm} $\forall k$. In addition, as the electrodes are often extremely long, an assumption that the leads are semi-infinite (i.e.,. $N_{L,R} = \infty$) has been considered as a good approximation and widely used in the literature. As explained above, only the surface Green's functions $(G^0_{L,R})_{11}$ is needed to compute  $\Sigma_{L,R}$. Using Eq. (\ref{eq:A1}), we hence derive the following equation
\begin{eqnarray}
(G^0_{L,R})_{11} = [A^{L,R}_{11} - A^{L,R}_{12} (G^0_{L,R})_{11} A^{L,R}_{21}]^{-1}.
\label{eq:T4}
\end{eqnarray}
Actually, this surface-Green's function can be solved by using iterative recursive, eigenchannel decomposition or
mode matching methods \cite{Sancho1984,MacKinnon1985,Umerski1997,Rivas2003,Rocha2006,Sascha2017}. Once $(G^0_{L,R})_{11}$ (then,  self-energies $\Sigma_{L,R}$) are determined, the device Green's function $G_C$ can be computed by solving Eq.(\ref{eq:T2}). 

In the device, possible scatterings can be incorporated, in particular, by modeling the self-energy $\Sigma_S$ for electron-phonon interactions, the potential energies $U$ for the effects of gate voltages or charge impurities, or by adjusting the Hamiltonian for computing the structural defects. In general, the matrix $A_C$ in Eq.(\ref{eq:T2}), therefore, is written as $A_C = E + i0^+ - H_C - U - \Sigma$ with $\Sigma = \Sigma_L + \Sigma_R + \Sigma_S$ in the considered two-terminal cases. Once all self-energies are computed and Eq. (\ref{eq:T2}) is solved, other Green's functions can be determined, i.e., $G^{\lessgtr}_C = G_C \Sigma^{\lessgtr}G^\dagger_C$ where $\Sigma^{\lessgtr}$ are lesser and greater self-energies, respectively. Here, $\Sigma^{<} = \Sigma_S^{<} + i\Gamma_L f_L + i\Gamma_R f_R$ and $\Sigma^{>} = \Sigma_S^{>} + i\Gamma_L (f_L-1) + i\Gamma_R (f_R-1)$, where $\Gamma_{L,R} = i(\Sigma_{L,R} - \Sigma^\dagger_{L,R})$ and $f_{L(R)}$ are the left (right) Fermi distribution functions. Finally, the electronic quantities can be computed using the following formulas 
\begin{itemize}
\item Local Density of States (LDOS) and of Charges (LDOC):  
\begin{equation}
    \mathrm{LDOS}(E,r_n) = - \frac{1}{\pi} \Im G_C (E,r_n,r_n), \hspace{0.5cm} \mathrm{LDOC}(r_n)= \int \mathrm{LDOS}(E,r_n) f (E) \, dE \nonumber
\end{equation}
\item Bond Current:
\begin{equation}
\mathrm{J}(r_n \to r_m) = 2\frac{e}{h} \int [H(r_n,r_m) G^<_C(r_m,r_n) - H(r_m,r_n) G^<_C (r_n,r_m)] \, dE \nonumber
\end{equation}
\item Total Transmission Function and Total Current:
\begin{equation}
\mathrm{T}(E) = \mathrm{Tr}[\Gamma_L G_C \Gamma_R G^\dagger_C], \hspace{0.5cm} \mathrm{J} = 2 \frac{e}{h} \int \mathrm{T}(E) [f_L(E)-f_R(E)]\, dE \nonumber
\end{equation}
\end{itemize}
Using the above-mentioned formulas and due to the tridiagonal-block form of the Hamiltonian matrix, it is not necessary to compute the full matrix of the Green's functions but some specific matrix blocks are sufficient to calculate the electronic and transport quantities. In particular, these specific blocks includes $G_{nn}$, $G_{n,n\pm1}$, $G_{n1}$, $G_{1n}$, $G_{nN}$ and $G_{Nn}$.

\subsection{Recursive calculations of the device Green's function}\label{sec3}

To compute the above-mentioned blocks of $G_C$, the recursive calculations have been developed \cite{Lewenkopf2013,Anantram2008,Do2014}. First, the right-connected Green's functions $g^R_{nn}$ are introduced, i.e., $g^R_{nn}$ is the Green's function of the \textit{n}th slice when it is decoupled with the slices in its left side. In particular, $g^R_{nn}$ are computed using the following equations
\begin{equation}
\begin{matrix}
\bullet & & & g^R_{NN} &=& (A_C)^{-1}_{NN} \\
\bullet & & & g^R_{nn} &=& [(A_C)_{nn} &-& (A_C)_{n,n+1}g^R_{n+1,n+1}(A_C)_{n+1,n}]^{-1}, \hspace{0.5cm} n = N-1, ..., 1.
\end{matrix} \label{eq:RD0}
\end{equation}
Based on Eqs. (\ref{eq:A1},\ref{eq:A4}), $(G_C)_{11} \equiv g^R_{11}$ is obtained as well as other diagonal blocks using 
\begin{equation}
(G_C)_{nn} = g^R_{nn} + g^R_{nn} (A_C)_{n,n-1} (G_C)_{n-1,n-1} (A_C)_{n-1,n} g^R_{nn}. \label{eq:RD1}
\end{equation}
Based on Eqs. (\ref{eq:A5},\ref{eq:A6}), off-diagonal blocks are computed as
\begin{eqnarray}
\begin{matrix}
(G_C)_{n-1,n} = - (G_C)_{n-1,n-1} (A_C)_{n-1,n} g^R_{n,n}, && (G_C)_{n,n-1} = - g^R_{n,n} (A_C)_{n,n-1} (G_C)_{n-1,n-1}
\\ 
(G_C)_{1n} = (G_C)_{11} \prod_{i=2}^{n} (-A_{i-1,i} g^R_{ii}), && (G_C)_{n1} = \left ( \prod_{i=n}^{2} (-g^R_{ii}A_{i,i-1}) \right ) (G_C)_{11}
\end{matrix} \label{eq:RD2}
\end{eqnarray}
Once these blocks are determined, all main quantities as transmission (therefore, current), local DOS (therefore, local charge density) and local current, etc., can be computed using the above-presented formulas.

Note that the computation cost to solve directly Eq.(\ref{eq:T2}) scales with $N_C^3$ where $N_C$ is the size of the device Hamiltonian $H_C$. Such a scaling makes calculations heavy, thus becoming unfeasible when modeling large devices. Using the recursive calculations, the computation cost however scales with $N_S M_{S}^3$ only where $N_S$ is the number of recursive slices and $M_S$ is their average size (i.e., $N_C = N_SM_S$). Calculations using the recursive techniques are thus optimized by a factor of $N^2_S$, compared to the full matrix ones, and the best optimization is obtained when the smallest recursive slices are computed.

\section{Large Superlattice-based Devices}\label{sec4}

To model large superlattice-based devices, the calculation of the device Green's functions can be optimized using Eqs. (\ref{eq:RD0}-\ref{eq:RD2}) and, as discussed, using the smallest recursive slices (i.e., not the large unit cell of superlattice). However, the partitioned slices used to derive Eq.(\ref{eq:T4}) have to be identical to the lead's periodic cells. Therefore, the calculations required to to compute the contact self-energies are still expensive and even become unfeasible when these periodic cells are large, i.e., the leads are also made of materials with large supercell. To solve this problem, two approaches allowing the use of recursive techniques to compute the contact self-energies are proposed herewith.

\subsection{Semi-infinite Lead Approach}\label{sec5}

\begin{figure}[ht]%
\centering
\includegraphics[width=0.7\textwidth]{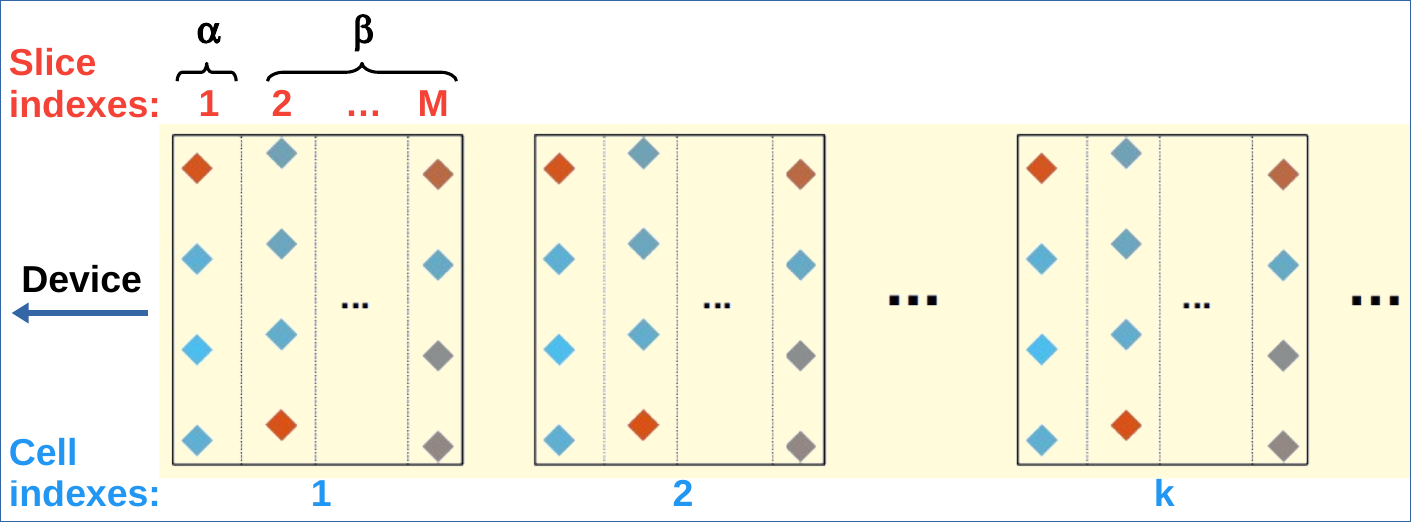}
\caption{Schematic atomistic model of the right semi-infinite-superlattice lead. The supercells consist of $M$ slices and are divided into two parts labelled ``$\alpha$" and ``$\beta$".}\label{figS02}
\end{figure}

We first consider the cases when the leads can be modeled as a semi-infinite and periodic superlattice. Note that in such a case, the self-energies in Eq.(\ref{eq:T2}) can be determined in term of the surface Green's function of atoms interacting directly with the device only, i.e., atoms in the part ``$\alpha$" of the 1st supercell illustrated in Fig.\ref{figS02}. 
Hence, the contact self-energy calculations can be optimized by recursively solving the Green's function $(G_{L,R})^{\alpha\alpha}_{11}$ (instead of the large matrix $(G_{L,R})_{11}$). Using the conventions represented in Fig.\ref{figS02}, the blocks of $A^{L,R}$ mentioned in Eq.(\ref{eq:T3}) are now formatted as
\begin{eqnarray}
A^{L,R}_{kk} = \begin{bmatrix}
A_{kk}^{\alpha\alpha} & A_{kk}^{\alpha\beta} \\ 
A_{kk}^{\beta\alpha} & A_{kk}^{\beta\beta}
\end{bmatrix}, \hspace{1cm} 
A^{L,R}_{k,k+1} = \begin{bmatrix}
0 & 0 \\ 
A_{k,k+1}^{\beta\alpha} & 0
\end{bmatrix}, \hspace{1cm} 
A^{L,R}_{k+1,k} = \begin{bmatrix}
0 & A_{k+1,k}^{\alpha\beta} \\ 
0 & 0
\end{bmatrix}.
\end{eqnarray}
Then, Eq. (\ref{eq:T3}) is rewritten as 
\begin{eqnarray}
\begin{bmatrix}
A_{11}^{\alpha\alpha} & A_{11}^{\alpha\beta} & 0 & \cdots & \cdots & \cdots \\ 
A_{11}^{\beta\alpha} & A_{11}^{\beta\beta} & A_{12}^{\beta\alpha} & 0 & \cdots & \cdots \\ 
0 & A_{21}^{\alpha\beta} & A_{22}^{\alpha\alpha} & A_{22}^{\alpha\beta} & 0 & \cdots \\ 
\cdots & 0 & A_{22}^{\beta\alpha} & A_{22}^{\beta\beta} & A_{23}^{\beta\alpha} & \cdots \\ 
\cdots & \cdots & 0 & A_{32}^{\alpha\beta} & A_{33}^{\alpha\alpha} & \cdots \\
\cdots & \cdots & \cdots & \cdots & \cdots & \cdots
\end{bmatrix} 
\begin{bmatrix}
G^{\alpha\alpha}_{11} & G^{\alpha\beta}_{11} & G^{\alpha\alpha}_{12} & G^{\alpha\beta}_{12} & G^{\alpha\alpha}_{13} & \cdots \\ 
G^{\beta\alpha}_{11} & G^{\beta\beta}_{11} & G^{\beta\alpha}_{12} & G^{\beta\beta}_{12} & G^{\beta\alpha}_{13} & \cdots \\ 
G^{\alpha\alpha}_{21} & G^{\alpha\beta}_{21} & G^{\alpha\alpha}_{22} & G^{\alpha\beta}_{22} & G^{\alpha\alpha}_{23} & \cdots \\ 
G^{\beta\alpha}_{21} & G^{\beta\beta}_{21} & G^{\beta\alpha}_{22} & G^{\beta\beta}_{22} & G^{\beta\alpha}_{23} & \cdots \\ 
G^{\alpha\alpha}_{31} & G^{\alpha\beta}_{31} & G^{\alpha\alpha}_{32} & G^{\alpha\beta}_{32} & G^{\alpha\alpha}_{33} & \cdots \\
\cdots & \cdots & \cdots & \cdots & \cdots & \cdots
\end{bmatrix} = \mathbb{I}. \label{eq:T5}
\end{eqnarray}
Since the considered leads are assumed periodic, $A^{L,R}_{kk} \equiv A^{L,R}_{11}$, $A^{L,R}_{k,k+1} \equiv A^{L,R}_{12}$, and $A^{L,R}_{k+1,k} \equiv A^{L,R}_{21}$ \hspace{0,1cm} $\forall \,\, k$.
Using the simplification suggested in (\ref{eq:A5b}), the matrix blocks $G^{\beta,:}_{k,:}$ and $G^{:,\beta}_{:,k}$ in Eq.(\ref{eq:T5}) can be eliminated and the following equation can be derived
\begin{eqnarray}
&& \begin{bmatrix}
\Pi_S & \Xi & 0 & \cdots & \cdots \\ 
\Xi^\dagger & \Pi & \Xi & 0 & \cdots \\ 
0 & \Xi^\dagger & \Pi & \Xi & \cdots \\
\cdots & 0 & \Xi^\dagger & \Pi & \cdots \\ 
\cdots & \cdots & \cdots & \cdots & \cdots 
\end{bmatrix}
\begin{bmatrix}
G^{\alpha\alpha}_{11}  & G^{\alpha\alpha}_{12} & G^{\alpha\alpha}_{13}  & G^{\alpha\alpha}_{14} & \cdots \\ 
G^{\alpha\alpha}_{21}  & G^{\alpha\alpha}_{22} & G^{\alpha\alpha}_{23}  & G^{\alpha\alpha}_{24} & \cdots \\ 
G^{\alpha\alpha}_{31}  & G^{\alpha\alpha}_{33} & G^{\alpha\alpha}_{33}  & G^{\alpha\alpha}_{34} & \cdots \\
G^{\alpha\alpha}_{41}  & G^{\alpha\alpha}_{43} & G^{\alpha\alpha}_{43}  & G^{\alpha\alpha}_{44} & \cdots \\
\cdots & \cdots & \cdots & \cdots & \cdots
\end{bmatrix} = \mathbb{I} \label{eq:T6} \\
\mathrm{with} \hspace{0.3cm} && \Pi_S = A_{11}^{\alpha\alpha} - A_{11}^{\alpha\beta} (A_{11}^{\beta\beta})^{-1} A_{11}^{\beta\alpha}, \,\, \Pi = \Pi_S - A_{21}^{\alpha\beta} (A_{11}^{\beta\beta})^{-1} A_{12}^{\beta\alpha}, \nonumber \\ 
&& {\color{white}{......}} \Xi = - A_{11}^{\alpha\beta} (A_{11}^{\beta\beta})^{-1} A_{12}^{\beta\alpha}, \,\, \Xi^\dagger = - A_{21}^{\alpha\beta} (A_{11}^{\beta\beta})^{-1} A_{11}^{\beta\alpha}. \nonumber
\end{eqnarray}
Based on Eq. (\ref{eq:A1}), $G^{\alpha\alpha}_{11}$ is computed by 
\begin{eqnarray}
G^{\alpha\alpha}_{11} = \begin{bmatrix}
\Pi_S - \Xi \widetilde{G}^{\alpha\alpha}_{22}
\Xi^\dagger \end{bmatrix}^{-1}, \label{eq:T7}
\end{eqnarray}
where $\widetilde{G}^{\alpha\alpha}_{22}$ is the solution of the equation
\begin{eqnarray}
&&\begin{bmatrix}
\Pi & \Xi & 0 & \cdots \\ 
\Xi^\dagger & \Pi & \Xi & \cdots \\ 
0 & \Xi^\dagger & \Pi & \cdots \\ 
\cdots & \cdots & \cdots & \cdots 
\end{bmatrix}\begin{bmatrix}
\widetilde{G}^{\alpha\alpha}_{22} & \widetilde{G}^{\alpha\alpha}_{23} & \widetilde{G}^{\alpha\alpha}_{24}  & \cdots \\ 
\widetilde{G}^{\alpha\alpha}_{32} & \widetilde{G}^{\alpha\alpha}_{33} & \widetilde{G}^{\alpha\alpha}_{34} & \cdots \\ 
\widetilde{G}^{\alpha\alpha}_{42} & \widetilde{G}^{\alpha\alpha}_{43} & \widetilde{G}^{\alpha\alpha}_{44} & \cdots \\
\cdots & \cdots & \cdots & \cdots
\end{bmatrix} = \mathbb{I}, \nonumber \\ \nonumber \\
\Rightarrow \,\,\,\,
&& \widetilde{G}^{\alpha\alpha}_{22} = [ \Pi - \Xi \widetilde{G}^{\alpha\alpha}_{22} \Xi^\dagger]^{-1}. \label{eq:T8}
\end{eqnarray}
Note that Eq.(\ref{eq:T8}) can be solved using the similar methods \cite{Sancho1984,MacKinnon1985,Umerski1997,Rivas2003,Rocha2006,Sascha2017} resolving Eq.(\ref{eq:T4}).

Thus, to compute $G^{\alpha\alpha}_{11}$ (i.e., solving Eqs.(\ref{eq:T7},\ref{eq:T8})), the matrices $\Pi_S$, $\Pi$, $\Xi$ and $\Xi^\dagger$ in Eq.(\ref{eq:T6}) must be determined. Actually, the matrix blocks $A_{11}^{\alpha\alpha}$, $A_{11}^{\alpha\beta}$ ($A_{11}^{\beta\alpha} \equiv  A_{11}^{\alpha\beta\dagger}$), $A_{11}^{\beta\beta}$, and $A_{12}^{\beta\alpha}$ ($A_{21}^{\alpha\beta} \equiv A_{12}^{\beta\alpha\dagger}$) are
\begin{eqnarray}
\begin{matrix}
A_{11}^{\alpha\alpha} = \widetilde{\mathcal{A}}_{11}, \,\, 
A_{11}^{\alpha\beta} = \begin{bmatrix}
\widetilde{\mathcal{A}}_{12} & 0 & 0 & \cdots  
\end{bmatrix}, \\
A_{12}^{\beta\alpha} = \begin{bmatrix}
0 \\ \cdots \\ 0 \\ \widetilde{\mathcal{A}}_{M1}  
\end{bmatrix}, \,\, A_{11}^{\beta\beta} = \begin{bmatrix}
\widetilde{\mathcal{A}}_{22} & \widetilde{\mathcal{A}}_{23} & 0 & \cdots & \\
\widetilde{\mathcal{A}}_{23} & \widetilde{\mathcal{A}}_{33} & \widetilde{\mathcal{A}}_{43} & 0 & \cdots \\ 
0 & \widetilde{\mathcal{A}}_{34} & \cdots &  &  \\ 
\cdots & 0 &  &  &  \\ 
 & \cdots &  &  & \widetilde{\mathcal{A}}_{MM}
\end{bmatrix}
\end{matrix}, %\nonumber
\end{eqnarray}
where $\widetilde{\mathcal{A}}_{nm} = (E + i0^+)\delta_{nm} - \widetilde{\mathcal{H}}_{nm}$ with $\widetilde{\mathcal{H}}_{nm}$ representing the Hamiltonian of the small slices in Fig.\ref{figS02} and their couplings. Therefore, we get
\begin{eqnarray}
\begin{matrix}
\Pi_S = {\color{white} s\,\,} \widetilde{\mathcal{A}}_{11} - \widetilde{\mathcal{A}}_{12} \widetilde{\mathcal{G}}_{22} \widetilde{\mathcal{A}}_{21}, \hspace{0,5cm}
\Pi {\color{white} s} = \Pi_S - \widetilde{\mathcal{A}}_{1M} \widetilde{\mathcal{G}}_{MM} \widetilde{\mathcal{A}}_{M1}, \\ 
\Xi {\color{white} s} =  - \widetilde{\mathcal{A}}_{12} \widetilde{\mathcal{G}}_{2M} \widetilde{\mathcal{A}}_{M1}, \hspace{1.3cm} \,\, 
\Xi^\dagger = - \widetilde{\mathcal{A}}_{1M} \widetilde{\mathcal{G}}_{M2} \widetilde{\mathcal{A}}_{21}, \,\hspace{0.9cm}
\end{matrix} \label{eq:T89a}
\end{eqnarray}
where $\widetilde{\mathcal{G}}$ is the solution of equation $A_{11}^{\beta\beta} \widetilde{\mathcal{G}} = \mathbb{I}$. Similar to the calculations presented Section \ref{sec3}, the Green's functions $\widetilde{g}^{R}_{nn}$ can be computed as 
$\widetilde{g}^R_{MM} = \widetilde{\mathcal{A}}^{-1}_{MM}$ and $\widetilde{g}^R_{mm} = [\widetilde{\mathcal{A}}_{mm} - \widetilde{\mathcal{A}}_{m,m+1} \widetilde{g}^R_{m+1,m+1} \widetilde{\mathcal{A}}_{m+1,m}]^{-1}$ with $m = M-1, ..., 2$. Then, $\widetilde{\mathcal{G}}_{22}$, $\widetilde{\mathcal{G}}_{2M}$, $\widetilde{\mathcal{G}}_{M2}$ and $\widetilde{\mathcal{G}}_{MM}$ in Eq.(\ref{eq:T89a}) are obtained by
\begin{eqnarray}
\begin{matrix}
\widetilde{\mathcal{G}}_{22} = \widetilde{g}^R_{22}, \,\,\,\, \widetilde{\mathcal{G}}_{mm} = \widetilde{g}^R_{mm} + \widetilde{g}^R_{mm} \widetilde{\mathcal{A}}_{m,m-1} \widetilde{\mathcal{G}}_{m-1,m-1} \widetilde{\mathcal{A}}_{m-1,m} \widetilde{g}^R_{mm}, {\color{white}{ssssssssss}}\\
\widetilde{\mathcal{G}}_{M2} = \left ( \prod_{i=M-1}^{2} (-\widetilde{g}^R_{i+1,i+1}\widetilde{\mathcal{A}}_{i+1,i}) \right ) \widetilde{\mathcal{G}}_{22}, \,\,\,\,
\widetilde{\mathcal{G}}_{2M} = \widetilde{\mathcal{G}}_{22} \prod_{i=2}^{M-1} (-\widetilde{\mathcal{A}}_{i,i+1}\widetilde{g}^R_{i+1,i+1})
\end{matrix}\label{eq:T89b}
\end{eqnarray}
Thus, after computing $\widetilde{\mathcal{G}}_{22}$, $\widetilde{\mathcal{G}}_{2M}$, $\widetilde{\mathcal{G}}_{M2}$ and $\widetilde{\mathcal{G}}_{MM}$, the matrices $\Pi_S$, $\Pi$, $\Xi$, and $\Xi^\dagger$ are determined using Eq.(\ref{eq:T89a}). Finally, the Green's functions $\widetilde{G}^{\alpha\alpha}_{22}$ and $G^{\alpha\alpha}_{11}$ are computed by solving Eq.(\ref{eq:T8}) and then Eq.(\ref{eq:T7}), respectively. 

Once again the advantage of the presented approach is to solve the small block $(G_{L,R})^{\alpha\alpha}_{11}$, instead of solving $(G_{L,R})_{11}$ of the whole large supercell. In general, the size of $(G_{L,R})^{\alpha\alpha}_{11}$ is much smaller than that of $(G_{L,R})_{11}$ in the case of large superlattices.

\subsection{Finite Lead Approach}\label{sec6}

\begin{figure}[b]%
\centering
\includegraphics[width=1\textwidth]{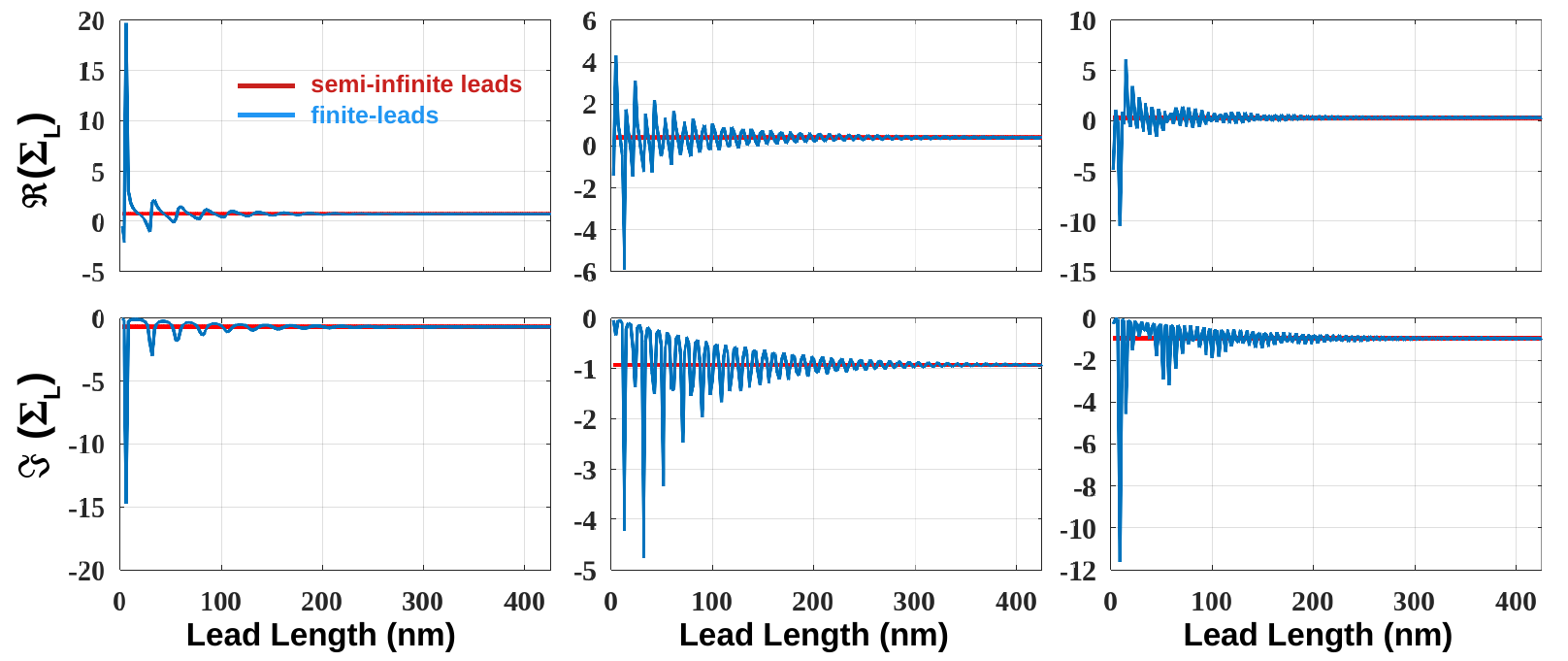}
\caption{Contact self-energies $\Sigma_L$ computed at three different energies (0.1 eV, 0.2 eV, and 0.3 eV, from left-to-right, respectively) in a graphene monolayer device: finite-lead approach compared to semi-infinite one.}\label{figC01}
\end{figure}
The semi-infinite lead approach has been widely used in the literature, mostly because the leads are practically long and hence the corresponding calculations are much more expensive than solving the self-consistent Eq.(\ref{eq:T4}). In addition, taking into account scatterings in the device centre region and computing the device scaling effects have been demonstrated to be a good approach to clarify the transport properties and predict the device performance in many cases, e.g. as illustrated in Refs. \cite{Niquet2014,Nguyen2014b}. Therefore, the model when the leads are assumed semi-infinite and periodic has been shown to be a good approach \cite{Anantram2008,Lewenkopf2013,Do2014} in most device simulation studies. However, we emphasize that although they can be long and made of a high quality material, the leads in practice are finite and generally aperiodic. The latter is due to the presence of unavoidable defects, impurities, external fields, etc. Therefore, the finite-lead calculations can be helpful in some practical cases, for instance, as it will be illustrated in Section \ref{sec8}.

\begin{figure}[b]%
\centering
\includegraphics[width=0.99\textwidth]{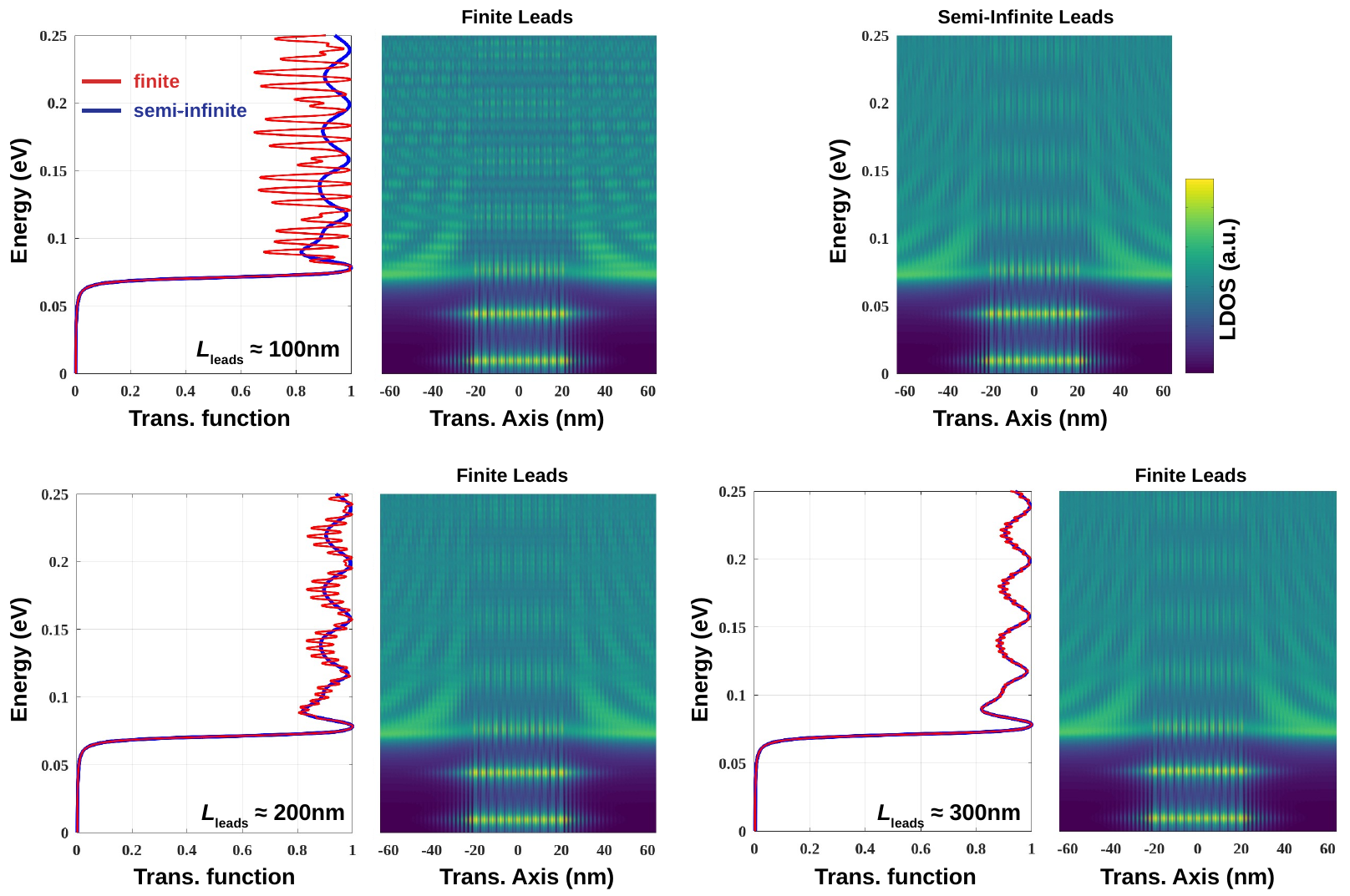}
\caption{Electronic transport (transmission function and corresponding LDOS) calculations performed for a single-potential-barrier monolayer graphene device, and computed with different lead lengths $L_{lead}$ using either the finite-lead or the semi-infinite-lead approaches for comparison.}\label{figR01a}
\end{figure}
In principle, in order to compute $G^{\alpha\alpha}_{11}$, it is not always necessary to solve the self-consistent equations Eq.(\ref{eq:T4}) or Eq.(\ref{eq:T8}). Indeed, using a finite-leads model, the surface Green's function $G^{\alpha\alpha}_{11}$ can be computed directly by the recursive calculations as presented in Eq.(\ref{eq:RD0}). Note that when the leads are sufficiently long (i.e., longer than the wavelength of charge carriers), these calculations can reach the semi-infinite limit as illustrated by the contact self-energies computed and presented in Fig.\ref{figC01} for a graphene monolayer-based device. In the other general cases, this approach can really take the finite-size effects of the leads into account in the calculations of contact self-energies. All these features are illustrated by the transport properties computed for a single potential barrier in monolayer graphene and presented in Fig.\ref{figR01a}. An important reason to use this approach to model large superlattice devices is that with the recursive techniques, it also allows the calculations of contact self-energies to be performed for small slices as presented in Fig.\ref{figS02} (instead of computing the whole large supercell as in Eq.(\ref{eq:T4})).

Thus, both the semi-infinite-lead and finite-lead approaches can be used in order to model large superlattice-based devices. However, each approach has its own advantages, so that they can complement each other. In particular, with the fast convergence scheme presented in Ref. \cite{Sancho1984} and improved calculations in Section \ref{sec5}, the semi-infinite lead approach could be a better method (i.e., a good approximation) for the devices with long and periodic leads. On the contrary, even though its convergence may be slower when imitating the limit of semi-infinite and periodic leads, the finite-lead approach can model more realistic devices, i.e., taking into account the effects of finite-size and aperiodic leads. Efficient applications of these two approaches will be illustrated and further discussed in Section \ref{sec8}.  
\section{Recursive Green's function of periodic systems}\label{sec7}

Computing the electronic band structure is another important task when modeling materials and their corresponding electronic devices. In principle, this is carried out by solving the eigenvalue equation of the Hamiltonian matrix and several numerical methods have been developed for this specific task. As it will be explained below, computing the Green's function of periodic (i.e., \textit{k-}dependent) Hamiltonians can be an alternative method and efficient in the case of large superlattices. This calculation is, in principle, expensive (even unfeasible) for large superlattices because of the large size of their Hamiltonian matrix. However, the use of recursive calculations can help to solve such a problem.
Besides the discussed computation capability, the Green's function calculations present another advantage compared to the diagonalization methods. Indeed, they allow to assess easily both the global and local electronic quantities as illustrated in Section \ref{sec2}. Consequently, different electronic aspects can be accurately clarified in details \cite{Nguyen2021}. However, taking into account the periodic boundary condition, recursive calculations presented in the previous sections are no longer valid and they hence have to be rewritten for periodic systems, as presented below. 

Modeling the electronic structure using the Green's function approach can be briefly explained as follow. Let us consider the \textit{k-}dependent Hamiltonian $H({\mathbf{k}})$ that possesses the following eigenvalues $\varepsilon_p ({\mathbf{k}})$ and  eigenwavefunctions $| \psi_p (\mathbf{k}) \rangle$. The Green's function of $H({\mathbf{k}})$ can be written as
\begin{eqnarray}
G(E,{\mathbf{k}}) = \sum_{p=1}^{N} \frac{\left | \psi_p ({\mathbf{k}}) \left \rangle  \right \langle \psi_p ({\mathbf{k}}) \right | } {E \,\,+\,\, i0^+ \,\,-\,\, \varepsilon_p ({\mathbf{k}})}. \label{eq:NEGEBands}
\end{eqnarray}
The Green's function clearly holds information about eigenvalues and eigenwavefunctions of the system and hence allows the extraction of some quantities to model its electronic properties. In particular, the \textit{k-}dependent local DOS computed from the diagonal elements of the Green's function (i.e., $\mathrm{LDOS}(\mathbf{k},E,\mathbf{r}) = -\Im G(\mathbf{k},E,\mathbf{r},\mathbf{r})/\pi$) allows for modeling (both globally and locally) the electronic structure of the system.

In order to go further, the recursive equations for the Green's function $G(\mathbf{k},E)$ of the discussing periodic systems has to be derived. First, one specific direction is selected (similar to the transport direction in Fig.\ref{figS01}) to partition the system into recursive slides. This direction should be parallel to one lattice vector of the superlattice. With the corresponding partitioned supercell, the equation of $G(\mathbf{k},E)$ can be written in the following form
\begin{eqnarray}
&& \begin{bmatrix}
A_{11} & A_{12} & 0 & \cdots & 0 & A_{1M} \\ 
A_{21} & A_{22} & A_{23}  &  &  & 0 \\ 
0 & A_{32} & \cdots & \cdots  &  & \cdots \\ 
\cdots &  & \cdots &  & & \\ 
0 &  &  &  & & \\ 
A_{M1} & 0 & \cdots &  & & A_{MM}
\end{bmatrix}\begin{bmatrix}
G_{11} & G_{12} & G_{13} & G_{14} & \cdots & G_{1M} \\ 
G_{21} & G_{22} & G_{23} & G_{24} & \cdots & \cdots \\ 
G_{31} & G_{32} & G_{33} & G_{34} & \cdots & \cdots \\ 
G_{41} & G_{42} & G_{43} & G_{44} & \cdots & \cdots \\ 
\cdots & \cdots & \cdots & \cdots & \cdots & \cdots \\ 
G_{M1} & \cdots & \cdots & \cdots & \cdots & G_{MM}
\end{bmatrix} = \mathbb{I} \label{eq:T9} \\ \nonumber \\
\mathrm{with} \,\, && A(\mathbf{k},E) = E + i0^+ - H(\mathbf{k}) \nonumber
\end{eqnarray}
where $A_{1M}$ and $A_{M1}$ blocks actually model the periodic boundary conditions. Here, the matrix blocks $H_{nm}$ (i.e., $A_{nm} = (E + i0^+)\delta_{nm} - H_{nm}$) represent the Hamiltonian of the small recursive slices (similar to those illustrated in Fig.\ref{figS02}) and their couplings. The recursive equations for diagonal blocks $G_{mm}$ are derived as follows. We first rewrite Eq.(\ref{eq:T9}) as
\begin{eqnarray}
&& A_{1:M-1,1:M-1} G_{1:M-1,1:M-1} = \mathbb{I} - A_{1:M-1,M} G_{M,1:M-1}, 
\label{eq:T12} \\
&& \mathrm{where} \,\,\,\,
A_{1:M-1,M} = [
\begin{matrix}
A_{1M} & 0 & 0 & \cdots & 0 & A_{M-1,M} 
\end{matrix}]^T. \nonumber
\end{eqnarray}
Note that $A_{1:M-1,1:M-1}$ is a tridiagonal-block matrix and hence $G_{mm}$ ($m=1,...,M-1$) can be recursively computed in terms of the blocks $G_{M,1:M-1}$, in a similar way as presented in the previous sections. Solving Eq.(\ref{eq:T12}), we indeed obtain
\begin{eqnarray}
G_{mm} = g^{LR}_{mm} + g^{LR}_{mm} (u^L_{mM}+u^R_{mM})G_{Mm}, \label{eq:Gmm}
\end{eqnarray}
where $g^{LR}_{mm} = [A_{mm} - A_{m,m-1}\overline{g}^L_{m-1,m-1}A_{m-1,m} - A_{m,m+1}\overline{g}^R_{m+1,m+1}A_{m+1,m}]^{-1}$ for $m = 2,...,M-1$, whereas 
$g^{LR}_{M-1,M-1} \equiv \overline{g}^L_{M-1,M-1}$ and $g^{LR}_{11} \equiv \overline{g}^R_{11}$. The Green's functions $\overline{g}^L_{mm}$ and $\overline{g}^R_{mm}$ are computed as
\begin{eqnarray}
&& {\color{white}{ssss}} \overline{g}^L_{11} = A^{-1}_{11}, \,\,\, \overline{g}^L_{mm} = [A_{mm} - A_{m,m-1}\overline{g}^L_{m-1,m-1}A_{m-1,m}]^{-1}, \label{eq:recurleft} \\
&& {\color{white}{ssss}} \overline{g}^R_{M-1,M-1} = A^{-1}_{M-1,M-1}, \,\,\, \overline{g}^R_{mm} = [A_{mm} - A_{m,m+1}\overline{g}^R_{m+1,m+1}A_{m+1,m}]^{-1}, \label{eq:recurright} 
\end{eqnarray}
and $\begin{matrix}
u^L_{mM} = - \left ( \prod_{i=m}^{2} (-A_{i,i-1}\overline{g}^L_{i-1,i-1}) \right )A_{1M}, \,\,
u^R_{mM} = - \left ( \prod_{i=m}^{M-2} (-A_{i,i+1}\overline{g}^R_{i+1,i+1}) \right )A_{M-1,M}
\end{matrix}$.
To compute $G_{Mm}$ (and $G_{MM}$ as well), the equation $G A = \mathbb{I}$ (particularly, \\ $\sum_{m=1}^{M} G_{Mm} A_{mn} = \mathbb{I} \delta_{n,M}$) is solved and we obtain
\begin{eqnarray}
\begin{matrix}
G_{MM} = [A_{MM} + z^L_{M1}\overline{g}^L_{11}A_{1M} + z^L_{M,M-1}\overline{g}^L_{M-1,M-1}A_{M-1,M}]^{-1},
\\ 
\mathrm{and} \,\,\, G_{Mm} = G_{MM}z^L_{Mm}\overline{g}^L_{mm}. {\color{white}{sssssssssssssssssssssssssssssssssssss}}
\end{matrix} \label{eq:GMm}
\end{eqnarray}
The matrix blocks $z^L_{Mm}$ are determined in terms of $\overline{g}^L_{mm}$ as
\begin{eqnarray}
z^L_{M,M-1} &=& u^L_{M,M-1} - A_{M,M-1}, \,\,\, z^L_{Mm} = u^L_{Mm} -  z^L_{M,m+1} \overline{g}^L_{m+1,m+1} A_{m+1,m}, \\
\mathrm{where} {\color{white}{aa}} u^L_{Mm} &=& \begin{matrix} - A_{M1} \prod_{i=2}^{m} (-\overline{g}^L_{i-1,i-1}A_{i-1,i}) 
\end{matrix}. \nonumber
\end{eqnarray}

Thus, all diagonal blocks of the Green's function (accordingly, its diagonal elements and then $\mathrm{LDOS}(\mathbf{k},E,\mathbf{r})$) can be computed, allowing to model the electronic bands, bands projected on specific orbitals/atoms (i.e., projected bands), total DOS and local DOS of the periodic systems (see in Ref. \cite{Nguyen2021} and as illustrated in the next section). As mentioned above, these calculations could be an alternative and efficient method, compared to the existing conventional diagonalization techniques, for investigating large superlattices. Note additionally that to model the electronic bands only, it is not necessary to compute all diagonal elements of the Green's function. In particular, as illustrated by Eq.(\ref{eq:NEGEBands}), the presence of an eigenvalue $\varepsilon_p ({\mathbf{k}})$ corresponds to a singularity in LDOS$(\mathbf{k},E)$ at $E \equiv \varepsilon_p ({\mathbf{k}})$. In addition, this singularity is generally obtained for all recursive slices. Therefore, any single block $G_{mm}$ computed using simple calculations presented in Appendix \ref{AppC} can be used to model the electronic bands. Actually, this method is even more convenient and cheaper than the above-presented calculations.

\section{Applications and discussions}\label{sec8}

Aiming to discuss the above-presented optimized calculations, twisted bilayer graphene (TBG) superlattices are now considered as an example to demonstrate the efficiency of the technique. These 2D materials are indeed typical superlattices, whose periodic length increases when reducing the twist angle $\theta$ and can reach tens to hundreds of nanometers, corresponding to a huge number (i.e., $> 10^4$) of atoms in their supercell. Importantly, the most interesting electronic features in TBGs have been observed in such a large size regime \cite{Nguyen2021,Gadelha2021,Nguyen2022} and hence our optimized calculations can be helpful for modeling their electronic devices. In addition, we will demonstrate that the recursive Green's function method can not only model the transport properties but can also be extended to compute other electronic aspects, e.g., to analyze the electronic structure, to compute magnetic field effects, etc. In the present section, the electronic properties of TBG superlattices are computed using the first-principles-enriched tight-binding Hamiltonian described in Ref. \cite{Nguyen2022}.

\subsection{Electronic properties of large superlattices}\label{sec10}

\begin{figure}[b]%
\centering
\includegraphics[width=0.8\textwidth]{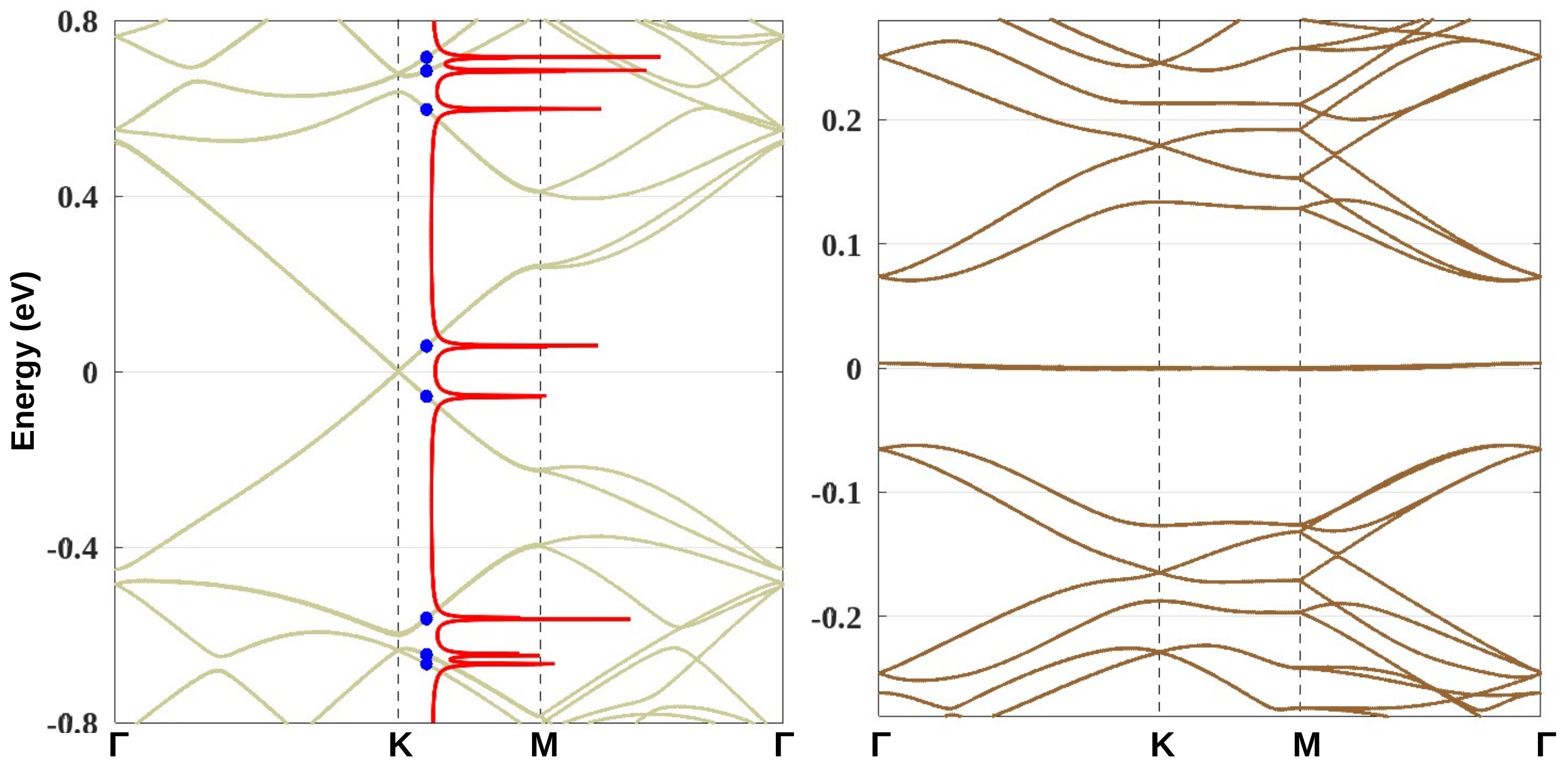}
\caption{Electronic structure of twisted bilayer graphene superlattices: $\theta \simeq 3.89^\circ$ (left) and $1.08^\circ$ (right). The electronic bands are modelled by local DOS computed using the Green's function techniques presented in Section \ref{sec7}. On the left, the local DOS and eigen-energies (marked by the blue circles) at a given momentum are superimposed, demonstrating the method to extract the electronic bands (see main text).}\label{figR10}
\end{figure}

In this subsection, the calculations described in Section \ref{sec7} are performed to model the electronic properties of TBG superlattices. In Fig. \ref{figR10}, the $k$-dependent local density of states $\mathrm{LDOS} (\mathbf{k},E,\mathbf{r})$ is computed and used to extract the electronic bandstructure for TBGs with two different twist-angles ($\theta \simeq 3.89^\circ$ and $1.08^\circ$). 
$\mathrm{LDOS} (\mathbf{k},E,\mathbf{r})$ at a given $k$-point marked by the blue circles is also superimposed, depicting the modeling technique. As explained above, $\mathrm{LDOS} (\mathbf{k},E,\mathbf{r})$ indeed presents high peaks when $E$ equals to any eigenvalue $\varepsilon_p ({\mathbf{k}})$ of $H(\mathbf{k})$. On this basis, the LDOS's peaks can be used to create the electronic bands as presented. In particular, the electronic features reported in the literature as Fermi velocity renormalization and low-energy van Hove singularities \cite{Trambly2010,Nguyen2015} are observed at $\theta \simeq 3.89^\circ$, while the flat bands are accurately reproduced at the magic angle $\theta \simeq 1.08^\circ$ \cite{MacDonald2011,Cao2018,Nguyen2021}. 

\begin{figure}[b]%
\centering
\includegraphics[width=0.95\textwidth]{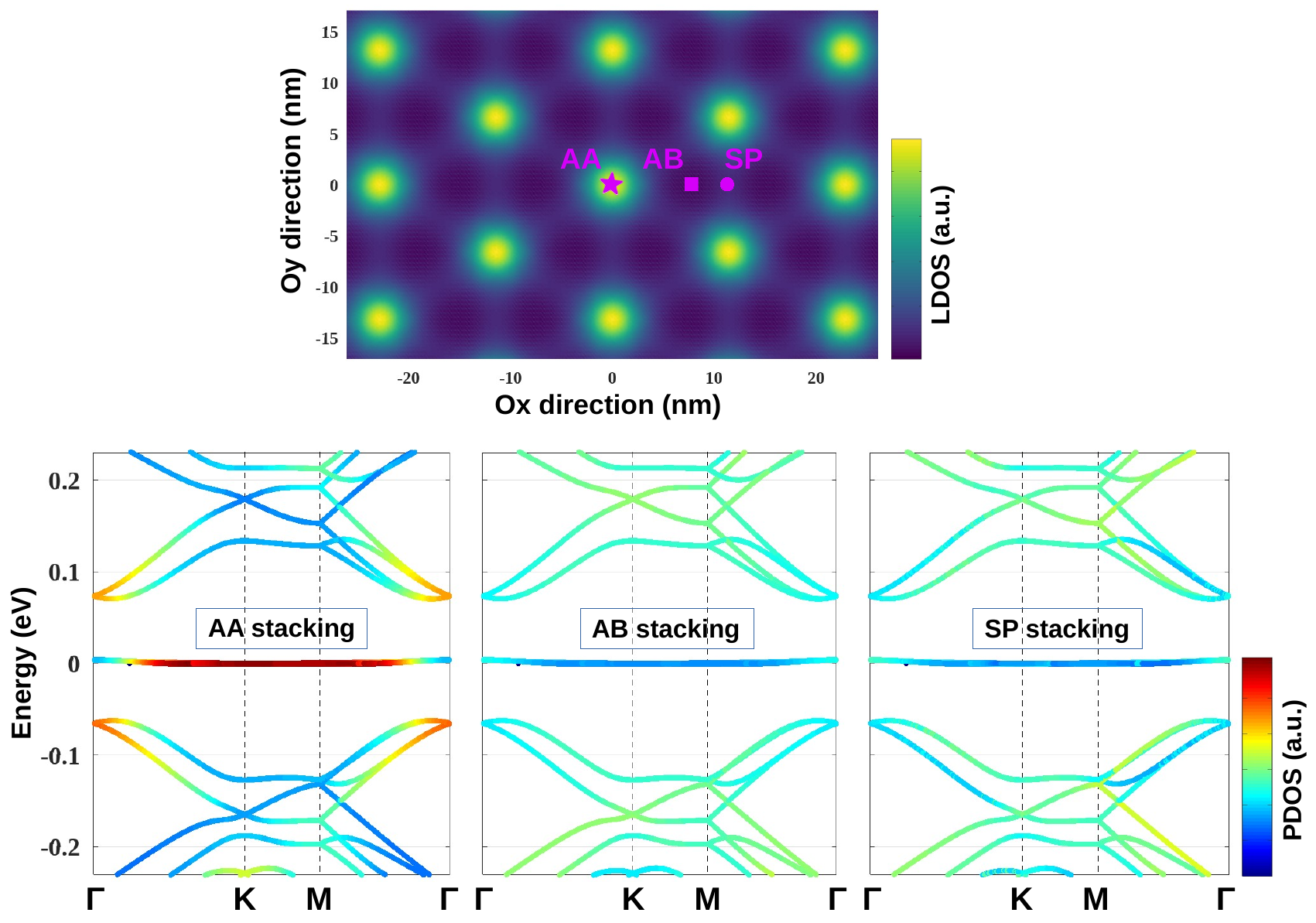}
\caption{Local densities of states at zero energy in twisted bilayer graphene superlattice with $\theta \simeq 1.08^\circ$ (top) and its corresponding electronic bands projected on different stacking regions (bottom), also marked in the top panel as AA-, AB-, SP-regions.}\label{figR11}
\end{figure}

In these calculations, the supercell is actually partitioned into $M = 14$ and 52 slices for $\theta \simeq 3.89^\circ$ and $1.08^\circ$, respectively. Note that the recursive calculation cost scales with $M$, whereas the standard diagonalization scales with $M^3$, i.e., the computation cost is thus reduced by a factor that scales with $M^2$. For instance, the reduction factors of computation time about 23 and 307 are obtained in the two previously considered cases $\theta \simeq 3.89^\circ$ and $1.08^\circ$, respectively. This optimization scheme is also discussed in Fig.\ref{figR02} in the next section. With this impressive reduction of the computation cost, the present recursive Green's functions technique can be used to compute much larger superlattices, as illustrated in Ref. \cite{Nguyen2021} for very small-angle TBG superlattices (e.g., $\theta \sim 0.2^\circ$ and accordingly $N_a \simeq$ 270 000 atoms in the supercell). In principle, these recursive calculations are still feasible for large superlattices with $N_a \sim 10^6$ (i.e., periodic length $\sim$ 10$^2$ \textit{nm}) since the number of atoms in recursive slices can still be small, i.e., in the order of $10^3$.

The advantages of these recursive calculations also include their capabilities to compute easily the local electronic quantities. This is indeed illustrated by local DOS and electronic bands projected to different stacking regions in the TBG superlattice at the magic angle presented in Fig.\ref{figR11}. More examples related to these types of calculations can also be found in Ref. \cite{Nguyen2021}, where several quantities have been computed in order to deeply clarify the local electronic properties of large TBG superlattices \cite{Nguyen2021,Gadelha2021}, which are not observed in conventional graphene.

\subsection{Electronic transport in large superlattice devices}\label{sec9}

\begin{figure}[t]%
\centering
\includegraphics[width=0.83\textwidth]{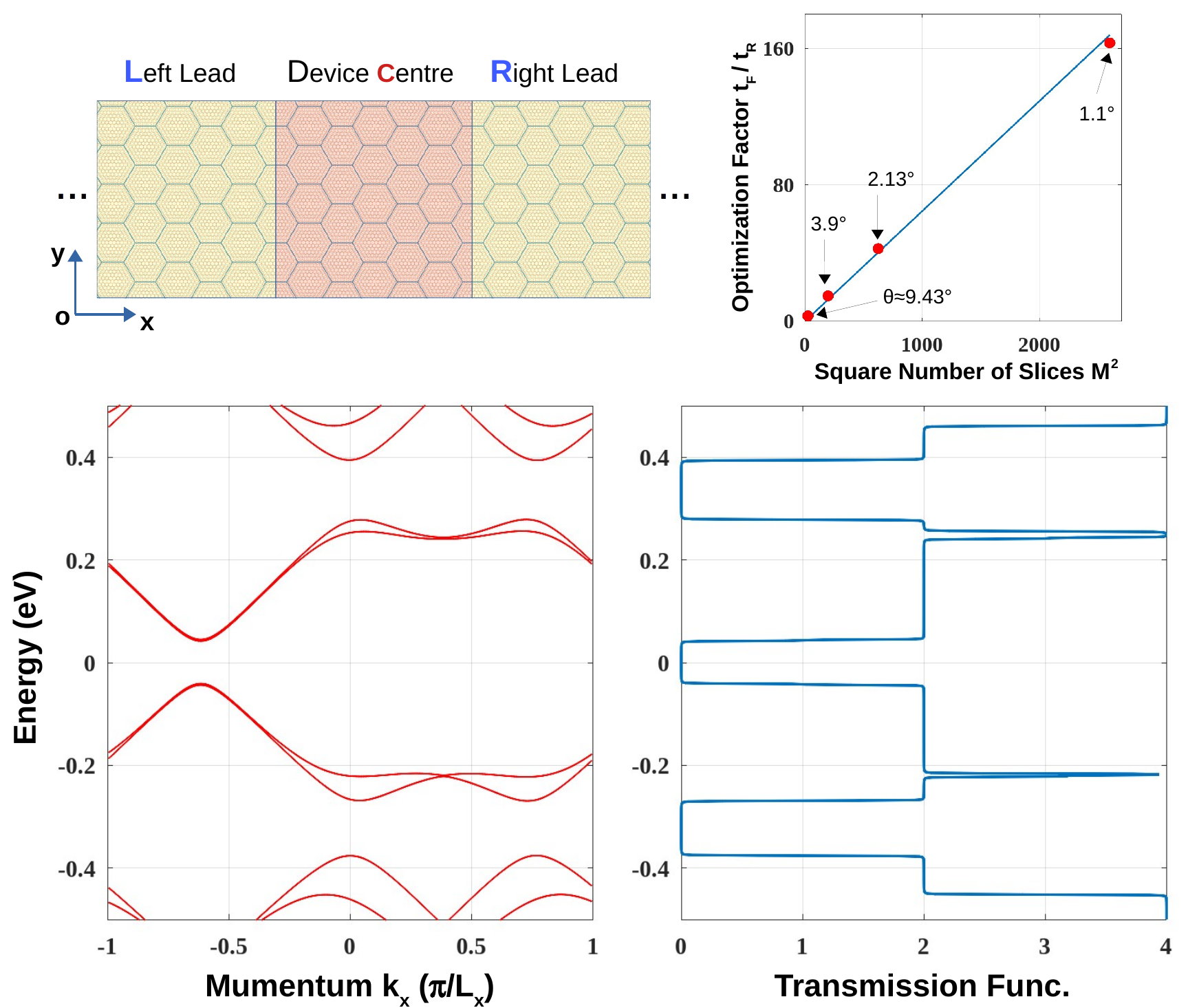}
\caption{Ballistic transport across a twisted ($\theta \simeq 3.89^\circ$) bilayer graphene superlattice. The computed channel is a 2D lattice and hence the periodic boundary condition is applied along the Oy axis (top-left panel). On the bottom panel, the electronic bands and the transmission function (computed for $k_y = 0.7667\,\pi/L_y$ where $L_y$ is the periodic length along the Oy direction), are presented. On the top-right panel, the ratio $t_F/t_R$ (computed at $E = 0.3\,eV$ for twisted superlattices at different twist angles) is presented, where $t_{F,R}$ are the elapsed times for computing the contact self-energies by the standard and by our recursive techniques, respectively.}\label{figR02}
\end{figure}

Let's now compute the electronic transport in TBG devices as schematized on the top-left panel of Fig.\ref{figR02}, using the calculations described in Section \ref{sec5}. The considered conducting channel is a 2D TBG sheet with periodicity along the Oy axis and the ballistic transport is computed along the Ox direction. 

In order to evaluate the efficiency of the newly developed calculations, the elapsed time for computing the lead Green's functions (accordingly, contact self-energies) is compared to the one based on the standard approach (see the results on top-right panel of Fig. \ref{figR02}). Actually, $t_R$ scales with the number $M$ of recursive slices in the supercell while $t_F \propto M^3$, i.e., the optimization factor $t_F / t_R \propto M^2$ as seen. Consequently, a very significant optimization can be observed, i.e., $t_F/t_S$ can reach the order of $\gtrsim 10^2$ for the small ($\lesssim 1.1^\circ$) twist angles (i.e., large TBG superlattices). Finally, the transmission function is accurately computed, which is confirmed by its perfect agreement with the ballistic channels identified by the electronic bands of the system (see bottom panels of Fig.\ref{figR02}).

\begin{figure}[t]%
\centering
\includegraphics[width=0.99\textwidth]{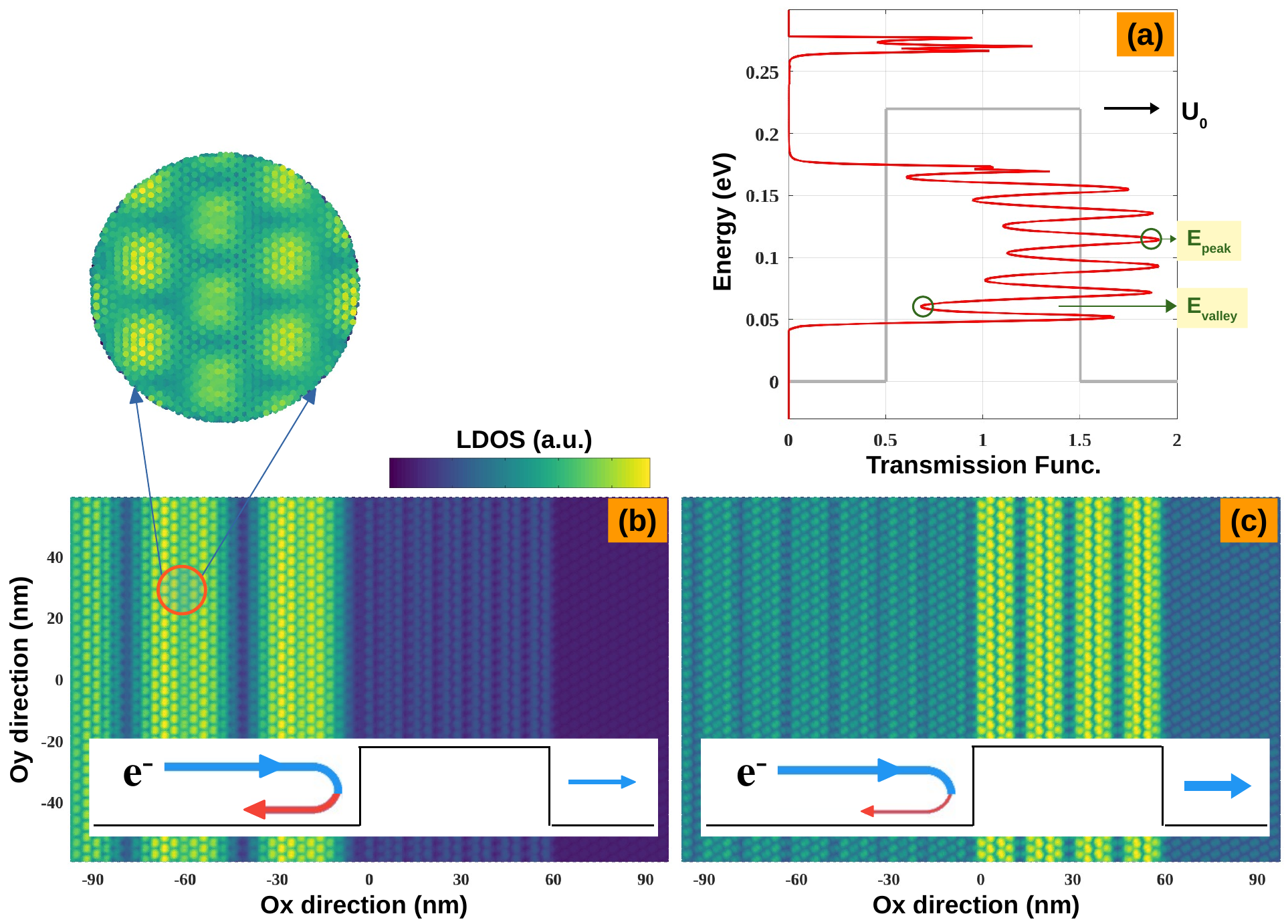}
\caption{Fabry-Pérot resonances in the transmission (a) through a potential barrier in a twisted ($\theta \simeq 3.89^\circ$) bilayer graphene device considered in Fig.\ref{figR02}. On the bottom panels, the left-injected local DOS computed at two energies $E_{valley}$ and $E_{peak}$ (as marked in (a)) are represented in (b) and (c), respectively.} \label{figR03}
\end{figure}

In the next step, the resonant transport through a single-potential-barrier in the TBG device (see Fig. \ref{figR03}) is computed. The presence of a potential barrier (practically created by the gate voltages \cite{Brun2020}) can induce scatterings and quantum confinement inside the barrier region. The latter have been shown to result in Fabry–Pérot interference and consequently in the observation of resonant transport features \cite{Young2009,Rickhaus2013,Brun2020} in monolayer graphene devices. Indeed, the resonant transport is similarly observed in the considered TBG device as illustrated in Fig. \ref{figR03}(a). To clarify the properties of these resonant features, LDOS along the transport direction could be helpful. In particular, the left-injected (similarly, right-injected) LDOS allows for imaging the electron propagation through the device \cite{Datta2000,Nguyen2020} and is computed by $\mathrm{LDOS}_{L} = G\Gamma_{L}G^{\dagger}/2\pi$. For this calculation, the Green's function blocks $G_{n1}$ must be computed, particularly, using Eq.(\ref{eq:RD2}). The left-injected LDOS in the considered TBG device is presented in Fig. \ref{figR03}(b,c) for two energies $E_{valley}$ and $E_{peak}$ (as marked in Fig. \ref{figR03}(a)), respectively. Obviously, the electron wave at $E = E_{valley}$ decays strongly when crossing the barrier, thus explaining the off-resonant transmission. At $E = E_{peak}$, the decay is relatively weak and accordingly the resonant transmission is observed. The computed LDOS pictures also allows imaging in a very accurate way the signatures of Moiré superlattice interactions (see the zoom-in image as an inset of Fig.\ref{figR03}(b)).

Besides the simple examples presented here, the recursive Green's function method can also conveniently compute the spatial bond current \cite{Zhang2019,Brun2019}, which has been shown to be very helpful for interpreting quantum transport phenomena (for instance, quantum interference effects). In addition, note that the most important capability of the method is to compute the transport performance of nanoscale electronic devices \cite{Anantram2008}, when implemented in the device simulation codes \cite{TB-SIM,NANOTCAD} and taking the interplay between the quantum confinement and scattering effects \cite{Nguyen2013,Nguyen2014b,Niquet2014,Mathieu2009} into account.

\subsection{Superlattices under magnetic field}\label{sec11}

Magnetic field effects inducing various fascinating quantum phenomena are an important aspect in nanophysics. Therefore, developing calculations to investigate the electronic properties in crystals under magnetic field is very desirable. However, there is one important obstacle: the presence of a magnetic field (often modeled by a vector potential) precludes the periodicity of the Hamiltonian and therefore the electronic structure calculations are generally very expensive. The periodic Landau gauge has hence been considered to cope such problem \cite{Nemec2007,Hasegawa2013,Moon2012,Moon2020,Wu2021}. 
However, this approach still presents two issues of principle: (i) the value of computed magnetic fields is discrete and (ii) the size of periodic cells is inversely proportional to the magnitude of magnetic field. The latter leads to very expensive calculations for the small magnetic fields ($\sim$ a few Tesla). 
\begin{figure}[t]%
\centering
\includegraphics[width=0.83\textwidth]{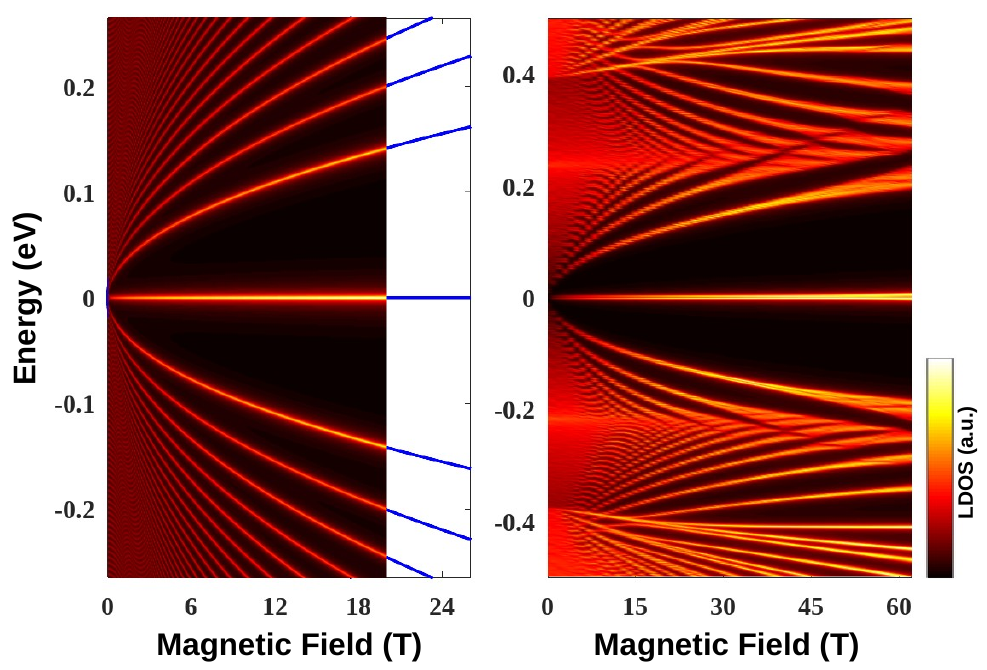}
\caption{Electronic spectra in monolayer graphene (left) and twisted ($\theta \simeq 3.89^\circ$) bilayer graphene superlattice (right) under magnetic field. In the left panel, the well-known Landau spectrum of monolayer graphene $E_n = \mathrm{sgn}(n)\sqrt{2e\hbar v^2_F \left | n \right | B}$ \cite{Yin2017} (blue curves) is superimposed to demonstrate the accuracy of the technique.}\label{figR04}
\end{figure}

Here, an alternative technique is proposed based on the finite-lead approach described in Section \ref{sec6} in order to solve the previously discussed issues. This proposal is actually based on an idea that the 2D lattice is essentially the infinite-size limit of a 1D lattice. Therefore, considering the device channel that is infinite along the vertical direction and computing the electronic spectrum of the finite-lead device when the leads are sufficiently long can be a good approach to mimic the effects of the magnetic field in 2D system. Note that in these calculations, the finite-size confinement and edge scattering effects (due to the finite size of the system along the transport axis) are unavoidable. However, as the leads are sufficiently long (i.e., longer than the wavelength of charge carriers), these effects could be negligible and the obtained results can really reach their 2D limit as already demonstrated in Section \ref{sec6}.

\begin{figure}[t]%
\centering
\includegraphics[width=0.77\textwidth]{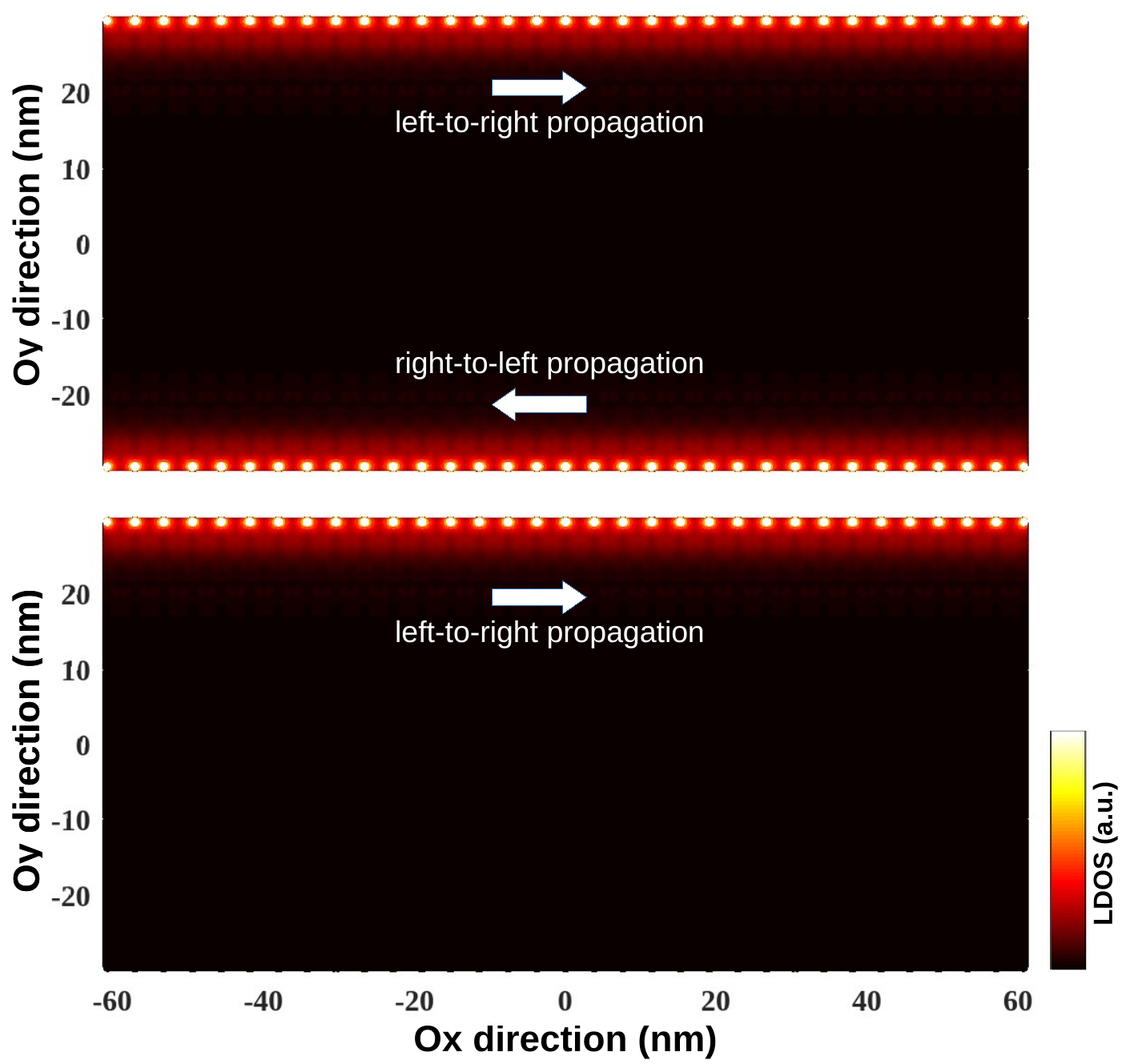}
\caption{Quantum Hall edge states predicted in a twisted ($\theta \simeq 3.89^\circ$) bilayer graphene nanoribbon ($\sim$ 60 nm wide). Calculations are performed at $B = 10\,T$ and $E = 60$ meV. Total and left-injected LDOS are presented on top and bottom, respectively.}
\label{figR05}
\end{figure}

The discussions above are illustrated by the electronic spectra presented in Fig.~\ref{figR04}. Here, the calculated systems are infinite and periodic along the Oy axis and the Gauge field $\vec{A} = \left ( 0, Bx, 0 \right )$ is chosen. Consequently, the Hamiltonian is still periodic along the Oy direction but aperiodic along the transport one (i.e., Ox axis). Indeed, the present calculations compute quite accurately the electronic spectra of monolayer and twisted bilayer graphene systems under out-of plane magnetic fields $B$. In particular, the Landau quantization $E_n = \mathrm{sgn}(n)\sqrt{2e\hbar v^2_F \left | n \right | B}$ \cite{Yin2017} (see  the blue curves superimposed in the left panel of Fig. \ref{figR04}) in the monolayer graphene is accurately reproduced and simultaneously the Hofstadter butterfly spectrum \cite{Moon2012,Hejazi2019} is obtained in the TBG superlattice.
Note that compared to other calculations in the literature (mostly based on standard diagonalization methods \cite{Moon2012,Hejazi2019}), this Green's function techniques present two advantages: (i) it can compute continuous values of magnetic field and (ii) the size of the system to be calculated is independent of the magnitude of the field, i.e., does not increase when decreasing the field.

Finally, calculations to image the electron propagation under a magnetic field are very desirable for clarifying its effects (e.g., quantum Hall, interference, etc.). In particular, the formation of quantum Hall edge states can be investigated by computing large ribbons and accordingly $\vec{A} = \left (-By, 0, 0 \right )$ could be the best choice as calculations can benefit from the ribbon periodicity along the Ox axis. Since the leads are periodic, the semi-infinite-lead approach can be used and is faster than finite-lead calculations, i.e., the contact self-energies are faster converged, as discussed in Section \ref{sec4}. Quantum Hall edge states in a TBG ribbon are computed and presented in Fig. \ref{figR05}. Indeed, the formation of edge states, similar to those observed in monolayer graphene ribbons \cite{Nguyen2020}, is clearly demonstrated and the signature of Moiré superlattice interactions is also visualized. These results suggest that the presented calculations could also be extremely helpful to explore the interplay between the effects of magnetic field and superlattice interactions.

\section{Conclusion}

Although demonstrated to be very powerful for atomistic modeling of nanoscale devices, the recursive Green's function method reported in the literature was facing difficult issues when computing large superlattices, mainly due to the heavy or even unfeasible calculation of the contact self-energies. In this article, recursive equations solving the lead Green's functions (then, contact self-energies) are developed, based on either the semi-infinite or finite lead models. The derivation of these equations is actually obtained due to the fact that the device-to-lead couplings take place only at the lead's surface, i.e., between the device and its surface slice only. The developed techniques are shown to keep optimizing significantly the Green's function calculations when modeling large superlattice devices. In particular, the optimization factor is proportional to the square number of recursive slices and can reach the order of $> 10^2$ for small-angle twisted bilayer graphene. In addition, the recursive Green's functions for periodic systems are also developed. These calculations allow computing both the global and local electronic quantities and hence can deeply elucidate different aspects of the electronic structure of the system. Typical examples presented in the article demonstrate the usefulness of the method in modeling both the electronic properties of large superlattices and the transport properties of their electronic devices. 

\section*{Acknowledgments} We acknowledge financial support from the European Union’s Horizon 2020 Research Project and Innovation Program — Graphene Flagship Core3 (N$^{\circ}$ 881603), from the Flag-Era JTC projects “TATTOOS” (N$^{\circ}$ R.8010.19) and “MINERVA” (N$^{\circ}$ R.8006.21), from the Pathfinder project “FLATS” (N$^{\circ}$ 101099139), from the F\'ed\'eration Wallonie-Bruxelles through the ARC Grant (N$^{\circ}$ 21/26-116) and  the EOS project “CONNECT” (N$^{\circ}$ 40007563), and from the Belgium F.R.S.-FNRS through the research project (N$^{\circ}$ T.029.22F). Computational resources have been provided by the CISM supercomputing facilities of UCLouvain and the C\'ECI consortium funded by F.R.S.-FNRS of Belgium (N$^{\circ}$ 2.5020.11).

\appendix
\section{Inverse of 2×2 and 3×3 block matrices} \label{secA} 

The inverse of a 2×2 block matrix $A$ is obtained by solving the equation
\begin{equation}
\begin{bmatrix}
A_{11} & A_{12} \\ 
A_{21} & A_{22}
\end{bmatrix}
\begin{bmatrix}
G_{11} & G_{12} \\ 
G_{21} & G_{22}
\end{bmatrix} = \mathbb{I} \,\,\,\,\,\, \mathrm{OR} \,\,\,\,\,\,
\begin{bmatrix}
G_{11} & G_{12} \\ 
G_{21} & G_{22}
\end{bmatrix}
\begin{bmatrix}
A_{11} & A_{12} \\ 
A_{21} & A_{22}
\end{bmatrix} = \mathbb{I}.
\end{equation}
The equations for matrix blocks $G_{11}$ and $G_{22}$ are hence derived as
\begin{eqnarray}
  \begin{bmatrix}
   A_{11} - A_{12}A_{22}^{-1}A_{21}
  \end{bmatrix} G_{11} &=& \mathbb{I}, \label{eq:A1} \\
  \begin{bmatrix}
   A_{22} - A_{21}A_{11}^{-1}A_{12}
  \end{bmatrix} G_{22} &=& \mathbb{I}. \label{eq:A2}
\end{eqnarray}
In addition, the following relationships are obtained
\begin{eqnarray}
G_{11} &=& A^{-1}_{11} + A^{-1}_{11} A_{12} G_{22} A_{21} A^{-1}_{11},\label{eq:A3} \\
G_{22} &=& A^{-1}_{22} + A^{-1}_{22} A_{21} G_{11} A_{12} A^{-1}_{22},\label{eq:A4} \\
G_{12} &=& - A^{-1}_{11} A_{12} G_{22} \equiv - G_{11} A_{12} A^{-1}_{22}, \label{eq:A5} \\
G_{21} &=& - A^{-1}_{22} A_{21} G_{11}. \label{eq:A6}
\end{eqnarray}

Similarly, solving the inverse equation of a 3×3 block matrix
\begin{eqnarray}
\begin{bmatrix}
A_{11} & A_{12} & A_{13} \\ 
A_{21} & A_{22} & A_{23} \\ 
A_{31} & A_{32} & A_{33}
\end{bmatrix}
\begin{bmatrix}
G_{11} & G_{12} & G_{13} \\ 
G_{21} & G_{22} & G_{23} \\ 
G_{31} & G_{32} & G_{33}
\end{bmatrix} = \mathbb{I},
\end{eqnarray}
and by eliminating blocks $G_{:,n}$ and $G_{n,:}$ ($n = 1,2,3$), we can derive the following equations :
\begin{eqnarray}
\begin{bmatrix}
A_{22}-A_{21}A^{-1}_{11}A_{12} & A_{23}-A_{21}A^{-1}_{11}A_{13} \\ 
A_{32}-A_{31}A^{-1}_{11}A_{12} & A_{33}-A_{31}A^{-1}_{11}A_{13}
\end{bmatrix}
\begin{bmatrix}
G_{22} & G_{23} \\
G_{32} & G_{33}
\end{bmatrix} = \mathbb{I}, \label{eq:A5a} \\
\begin{bmatrix}
A_{11}-A_{12}A^{-1}_{22}A_{21} & A_{13}-A_{12}A^{-1}_{22}A_{23} \\ 
A_{31}-A_{32}A^{-1}_{22}A_{21} & A_{33}-A_{32}A^{-1}_{22}A_{23}
\end{bmatrix}
\begin{bmatrix}
G_{11} & G_{13} \\
G_{31} & G_{33}
\end{bmatrix} = \mathbb{I}, \label{eq:A5b} \\
\begin{bmatrix}
A_{11}-A_{13}A^{-1}_{33}A_{31} & A_{12}-A_{13}A^{-1}_{33}A_{32} \\ 
A_{21}-A_{23}A^{-1}_{33}A_{31} & A_{22}-A_{23}A^{-1}_{33}A_{32}
\end{bmatrix}
\begin{bmatrix}
G_{11} & G_{12} \\
G_{21} & G_{22}
\end{bmatrix} = \mathbb{I}. \label{eq:A5c}
\end{eqnarray}

Note that in most text books and papers of the NEGF method, the recursive calculations have been derived using the Dyson's equations \cite{Lewenkopf2013,Anantram2008}. However, we show here that the derivation of these calculations can also be presented in a simpler way, i.e., based on the matrix algebras above and the tridiagonal-block form of the Hamiltonian matrix. Indeed, the right-connected Green's function $g^R_{nn}$ in Section \ref{sec3} is actually the solution of the equation $A_{n:N,n:N} g^R_{n:N,n:N} = \mathbb{I}$ and hence Eqs.(\ref{eq:RD0}) are obtained using Eq.(\ref{eq:A1}). By the definition, we obviously have $G_{11} \equiv g^R_{11}$. Then, Eqs.(\ref{eq:RD1},\ref{eq:RD2}) are obtained using Eqs.(\ref{eq:A4},\ref{eq:A5},\ref{eq:A6}).

%\clearpage
\section{Periodic systems: computing a single Green's function block} \label{AppC}

To compute a single block $G_{mm}$ of the Green's function in Eq.(\ref{eq:T9}), two following calculations could be used. Actually, these calculations are established to eliminate all other blocks of the Green's function, i.e., the equation of $G_{mm}$ is finally derived. 

In particular, to derive the equation for $G_{1:M-1,1:M-1}$, the blocks $G_{M,1:M}$ and $G_{1:M,M}$ are eliminated using Eq. (\ref{eq:A1})
\begin{eqnarray}
&&A^R_{1:M-1,1:M-1} G_{1:M-1,1:M-1} = \mathbb{I} \label{eq:C1} \\
&&A^R_{1:M-1,1:M-1} = A_{1:M-1,1:M-1} - A_{1:M-1,M}(A_{MM})^{-1}A_{M,1:M-1}. \nonumber
\end{eqnarray}
Here, $A^R_{11} = A_{11} - A_{1M}(A_{MM})^{-1}A_{M1}$, $A^R_{M-1,1} = - A_{M-1,M}(A_{MM})^{-1}A_{M1}$, $A^R_{1,M-1} = - A_{1M}(A_{MM})^{-1}A_{M,M-1}$, $A^R_{M-1,M-1} = A_{M-1,M-1} - A_{M-1,M}(A_{MM})^{-1}A_{M,M-1}$, and other blocks $A^R_{pq} \equiv A_{pq}$. Applying sequentially this calculation $M - m$ times, we derive the equation for the Green's function blocks $G_{1:m,1:m}$.

Similarly, the equation for $G_{2:M,2:M}$ is derived using Eq. (\ref{eq:A2})
\begin{eqnarray}
&&A^L_{2:M,2:M} G_{2:M,2:M} = \mathbb{I} \label{eq:C2} \\
&&A^L_{2:M,2:M} = A_{2:M,2:M} - A_{2:M,1}A^{-1}_{11}A_{1,2:M}. \nonumber
\end{eqnarray}
Here, $A^L_{22} = A_{22} - A_{21}A^{-1}_{11}A_{12}$, $A^L_{M2} = - A_{M1}A^{-1}_{11}A_{12}$, $A^L_{2M} = - A_{21}A^{-1}_{11}A_{1M}$, $A^L_{MM} = A_{MM} - A_{M1}A^{-1}_{11}A_{1M}$, and other blocks $A^L_{pq} \equiv A_{pq}$. Applying sequentially this calculation $m-1$ times, we finally derive the equation for the Green's function blocks $G_{m:M,m:M}$.

Thus, performing sequentially the calculations (\ref{eq:C1}) $M-m$ times and then the calculations (\ref{eq:C2}) $m-1$ times, the equation for a single bock $G_{mm}$ is obtained.

\bibliographystyle{unsrt}
\bibliography{Nguyen_biblio}

\begin{thebibliography}{10}

\bibitem{Granzner2006}
R.~Granzner, V.M. Polyakov, F.~Schwierz, M.~Kittler, R.J. Luyken, W.~Rösner, and M.~Städele.
\newblock Simulation of nanoscale mosfets using modified drift-diffusion and hydrodynamic models and comparison with monte carlo results.
\newblock {\em Microelectronic Engineering}, 83(2):241--246, 2006.

\bibitem{Fan2014}
Zheyong Fan, Andreas Uppstu, Topi Siro, and Ari Harju.
\newblock Efficient linear-scaling quantum transport calculations on graphics processing units and applications on electron transport in graphene.
\newblock {\em Computer Physics Communications}, 185(1):28--39, 2014.

\bibitem{Fan2021}
Zheyong Fan, José~H. Garcia, Aron~W. Cummings, Jose~Eduardo Barrios-Vargas, Michel Panhans, Ari Harju, Frank Ortmann, and Stephan Roche.
\newblock Linear scaling quantum transport methodologies.
\newblock {\em Physics Reports}, 903:1--69, 2021.
\newblock Linear scaling quantum transport methodologies.

\bibitem{Jacoboni1983}
Carlo Jacoboni and Lino Reggiani.
\newblock The monte carlo method for the solution of charge transport in semiconductors with applications to covalent materials.
\newblock {\em Rev. Mod. Phys.}, 55:645--705, 1983.

\bibitem{Hong2009}
S.-M. Hong and C.~Jungemann.
\newblock A fully coupled scheme for a boltzmann-poisson equation solver based on a spherical harmonics expansion.
\newblock {\em J. Comput. Electron.}, 8:225--241, 2009.

\bibitem{Datta1995}
Supriyo Datta.
\newblock {\em Electronic Transport in Mesoscopic Systems}.
\newblock Cambridge Studies in Semiconductor Physics and Microelectronic Engineering. Cambridge University Press, 1995.

\bibitem{Mingo2007}
W.~Zhang, T.~S. Fisher, and N.~Mingo.
\newblock The atomistic green's function method: An efficient simulation approach for nanoscale phonon transport.
\newblock {\em Numerical Heat Transfer, Part B: Fundamentals}, 51(4):333--349, 2007.

\bibitem{Pourfath2014}
Pourfath Mahdi.
\newblock {\em The Non-Equilibrium Green's Function Method for Nanoscale Device Simulation}.
\newblock Springer-Verlag, New York, 2014.

\bibitem{Lewenkopf2013}
Caio~H. Lewenkopf and Eduardo~R. Mucciolo.
\newblock The recursive green’s function method for graphene.
\newblock {\em Journal of Computational Electronics}, 12(2):203--231, 2013.

\bibitem{Niquet2014}
Yann-Michel Niquet, Viet-Hung Nguyen, François Triozon, Ivan Duchemin, Olivier Nier, and Denis Rideau.
\newblock Quantum calculations of the carrier mobility: Methodology, matthiessen's rule, and comparison with semi-classical approaches.
\newblock {\em Journal of Applied Physics}, 115(5):054512, 2014.

\bibitem{Lake1997}
Roger Lake, Gerhard Klimeck, R.~Chris Bowen, and Dejan Jovanovic.
\newblock Single and multiband modeling of quantum electron transport through layered semiconductor devices.
\newblock {\em Journal of Applied Physics}, 81(12):7845--7869, 1997.

\bibitem{Svizhenko2002}
A.~Svizhenko, M.~P. Anantram, T.~R. Govindan, B.~Biegel, and R.~Venugopal.
\newblock Two-dimensional quantum mechanical modeling of nanotransistors.
\newblock {\em Journal of Applied Physics}, 91(4):2343--2354, 2002.

\bibitem{Li2007}
S.~Li, S.~Ahmed, and E.~Darve.
\newblock Fast inverse using nested dissection for negf.
\newblock {\em J. Comput. Electron.}, 6:187--190, 2007.

\bibitem{Kazymyrenko2008}
K.~Kazymyrenko and X.~Waintal.
\newblock Knitting algorithm for calculating green functions in quantum systems.
\newblock {\em Phys. Rev. B}, 77:115119, 2008.

\bibitem{Anantram2008}
M.~P. Anantram, Mark~S. Lundstrom, and Dmitri~E. Nikonov.
\newblock Modeling of nanoscale devices.
\newblock {\em Proceedings of the IEEE}, 96(9):1511--1550, 2008.

\bibitem{Stephen2011}
Stephen Cauley, Mathieu Luisier, Venkataramanan Balakrishnan, Gerhard Klimeck, and Cheng-Kok Koh.
\newblock Distributed non-equilibrium green’s function algorithms for the simulation of nanoelectronic devices with scattering.
\newblock {\em Journal of Applied Physics}, 110(4):043713, 2011.

\bibitem{Do2014}
Van-Nam Do.
\newblock Non-equilibrium green function method: theory and application in simulation of nanometer electronic devices.
\newblock {\em Adv. Nat. Sci: Nanosci. Nanotechnol.}, 5(3):033001, 2014.

\bibitem{Thorgilsson2014}
G.~Thorgilsson, G.~Viktorsson, and S.I. Erlingsson.
\newblock Recursive green's function method for multi-terminal nanostructures.
\newblock {\em J. Comput. Phys.}, 261:256--266, 2014.

\bibitem{Zhang2019}
X.~W. Zhang and Y.~L. Liu.
\newblock Electronic transport and spatial current patterns of 2d electronic system: A recursive green’s function method study.
\newblock {\em AIP Advances}, 9(11):115209, 2019.

\bibitem{Mathieu2006}
Mathieu Luisier, Andreas Schenk, Wolfgang Fichtner, and Gerhard Klimeck.
\newblock Atomistic simulation of nanowires in the $s{p}^{3}{d}^{5}{s}^{*}$ tight-binding formalism: From boundary conditions to strain calculations.
\newblock {\em Phys. Rev. B}, 74:205323, 2006.

\bibitem{Bescond2007}
Marc Bescond, Nicolas Cavassilas, and Michel Lannoo.
\newblock Effective-mass approach for n-type semiconductor nanowire mosfets arbitrarily oriented.
\newblock {\em Nanotechnology}, 18(25):255201, 2007.

\bibitem{Mathieu2009}
Mathieu Luisier and Gerhard Klimeck.
\newblock Atomistic full-band simulations of silicon nanowire transistors: Effects of electron-phonon scattering.
\newblock {\em Phys. Rev. B}, 80:155430, 2009.

\bibitem{Mathieu2010}
Mathieu Luisier and Gerhard Klimeck.
\newblock Simulation of nanowire tunneling transistors: From the wentzel–kramers–brillouin approximation to full-band phonon-assisted tunneling.
\newblock {\em Journal of Applied Physics}, 107(8):084507, 2010.

\bibitem{Fiori2013}
Gianluca Fiori and Giuseppe Iannaccone.
\newblock Multiscale modeling for graphene-based nanoscale transistors.
\newblock {\em Proceedings of the IEEE}, 101(7):1653--1669, 2013.

\bibitem{Alarcon2013}
Alfonso Alarcon, Viet-Hung Nguyen, Salim Berrada, Damien Querlioz, Jérôme Saint-Martin, Arnaud Bournel, and Philippe Dollfus.
\newblock Pseudosaturation and negative differential conductance in graphene field-effect transistors.
\newblock {\em IEEE Transactions on Electron Devices}, 60:985--991, 2013.

\bibitem{Nguyen2013}
Viet-Hung Nguyen, François Triozon, Frédéric D.~R. Bonnet, and Yann-Michel Niquet.
\newblock Performances of strained nanowire devices: Ballistic versus scattering-limited currents.
\newblock {\em IEEE Transactions on Electron Devices}, 60, 2013.

\bibitem{Bescond2014}
Nicolas Cavassilas, Fabienne Michelini, and Marc Bescond.
\newblock Theoretical comparison of multiple quantum wells and thick-layer designs in ingan/gan solar cells.
\newblock {\em Applied Physics Letters}, 105(6):063903, 2014.

\bibitem{Nguyen2014b}
Viet-Hung Nguyen, Yann-Michel Niquet, François Triozon, Ivan Duchemin, Olivier Nier, and Denis Rideau.
\newblock Quantum modeling of the carrier mobility in fdsoi devices.
\newblock {\em IEEE Transactions on Electron Devices}, 61(9):3096--3102, 2014.

\bibitem{Bescond2017}
Nicolas Cavassilas, Yann Claveau, Marc Bescond, and Fabienne Michelini.
\newblock Quantum electronic transport in polarization-engineered gan/ingan/gan tunnel junctions.
\newblock {\em Applied Physics Letters}, 110(16):161106, 2017.

\bibitem{Zhang2017}
H~Zhang, N~Guan, V~Piazza, A~Kapoor, C~Bougerol, F~H Julien, A~V Babichev, N~Cavassilas, M~Bescond, F~Michelini, M~Foldyna, E~Gautier, C~Durand, J~Eymery, and M~Tchernycheva.
\newblock Comprehensive analyses of core–shell ingan/gan single nanowire photodiodes.
\newblock {\em Journal of Physics D: Applied Physics}, 50(48):484001, 2017.

\bibitem{Choukroun2019}
Jean Choukroun, Marco Pala, Shiang Fang, Efthimios Kaxiras, and Philippe Dollfus.
\newblock High performance tunnel field effect transistors based on in-plane transition metal dichalcogenide heterojunctions.
\newblock {\em Nanotechnology}, 30:025201, nov 2018.

\bibitem{Bescond2004}
M.~Bescond, J.L. Autran, D.~Munteanu, and M.~Lannoo.
\newblock Atomic-scale modeling of double-gate mosfets using a tight-binding green’s function formalism.
\newblock {\em Solid-State Electronics}, 48(4):567--574, 2004.

\bibitem{Antonio2007}
Antonio Martinez, Marc Bescond, John~R. Barker, Alexei Svizhenko, M.~P. Anantram, Campbell Millar, and Asen Asenov.
\newblock A self-consistent full 3-d real-space negf simulator for studying nonperturbative effects in nano-mosfets.
\newblock {\em IEEE Transactions on Electron Devices}, 54:2213--2222, 2007.

\bibitem{Groth2014}
Christoph~W Groth, Michael Wimmer, Anton~R Akhmerov, and Xavier Waintal.
\newblock Kwant: a software package for quantum transport.
\newblock {\em New Journal of Physics}, 16(6):063065, 2014.

\bibitem{Nguyen2018}
V.~Hung Nguyen and J.-C. Charlier.
\newblock Klein tunneling and electron optics in dirac-weyl fermion systems with tilted energy dispersion.
\newblock {\em Phys. Rev. B}, 97:235113, 2018.

\bibitem{Brun2019}
B.~Brun, N.~Moreau, S.~Somanchi, V.-H. Nguyen, K.~Watanabe, T.~Taniguchi, J.-C. Charlier, C.~Stampfer, and B.~Hackens.
\newblock Imaging dirac fermions flow through a circular veselago lens.
\newblock {\em Phys. Rev. B}, 100:041401, 2019.

\bibitem{Nguyen2020}
V.~Hung Nguyen and J.-C. Charlier.
\newblock Aharonov–bohm interferences in polycrystalline graphene.
\newblock {\em Nanoscale Adv.}, 2:256--263, 2020.

\bibitem{Ozaki2010}
Taisuke Ozaki, Kengo Nishio, and Hiori Kino.
\newblock Efficient implementation of the nonequilibrium green function method for electronic transport calculations.
\newblock {\em Phys. Rev. B}, 81:035116, 2010.

\bibitem{Papior2017}
Nick Papior, Nicolás Lorente, Thomas Frederiksen, Alberto García, and Mads Brandbyge.
\newblock Improvements on non-equilibrium and transport green function techniques: The next-generation transiesta.
\newblock {\em Comp. Phys. Comm.}, 212:8--24, 2017.

\bibitem{Zhang2007}
W.~Zhang, T.~S. Fisher, and N.~Mingo.
\newblock The atomistic green's function method: An efficient simulation approach for nanoscale phonon transport.
\newblock {\em Numerical Heat Transfer, Part B: Fundamentals}, 51(4):333--349, 2007.

\bibitem{Jinghua2009}
Jinghua Lan, Jian-Sheng Wang, Chee~Kwan Gan, and Sai~Kong Chin.
\newblock Edge effects on quantum thermal transport in graphene nanoribbons: Tight-binding calculations.
\newblock {\em Phys. Rev. B}, 79:115401, 2009.

\bibitem{Mazzamuto2011}
F.~Mazzamuto, V.~Hung~Nguyen, Y.~Apertet, C.~Ca\"er, C.~Chassat, J.~Saint-Martin, and P.~Dollfus.
\newblock Enhanced thermoelectric properties in graphene nanoribbons by resonant tunneling of electrons.
\newblock {\em Phys. Rev. B}, 83:235426, 2011.

\bibitem{Mazzamuto2012}
Fulvio Mazzamuto, Jerome Saint-Martin, Viet~Hung Nguyen, Christophe Chassat, and Philippe Dollfus.
\newblock Thermoelectric performance of disordered and nanostructured graphene ribbons using green’s function method.
\newblock {\em Journal of Computational Electronics}, 11:67--77, 2012.

\bibitem{Nguyen2014}
V.~Hung~Nguyen, M.~Chung Nguyen, Huy-Viet Nguyen, J.~Saint-Martin, and P.~Dollfus.
\newblock Enhanced thermoelectric figure of merit in vertical graphene junctions.
\newblock {\em Applied Physics Letters}, 105(13):133105, 2014.

\bibitem{Wang2014}
Jian-Sheng Wang, Bijay~Kumar Agarwalla, Huanan Li, and Juzar Thingna.
\newblock Nonequilibrium green’s function method for quantum thermal transport.
\newblock {\em Frontiers of Physics}, 9:673--679, 2014.

\bibitem{Drouvelis2006}
P.S. Drouvelis, P.~Schmelcher, and P.~Bastian.
\newblock Parallel implementation of the recursive green’s function method.
\newblock {\em Journal of Computational Physics}, 215(2):741--756, 2006.

\bibitem{Avouris2017}
Phaedon Avouris, Tony~F. Heinz, and Tony Low.
\newblock {\em 2D Materials: Properties and Devices}.
\newblock Cambridge University Press, 2017.

\bibitem{Ferrari2015}
Andrea~C. Ferrari and et~al.
\newblock Science and technology roadmap for graphene{,} related two-dimensional crystals{,} and hybrid systems.
\newblock {\em Nanoscale}, 7:4598--4810, 2015.

\bibitem{Geim2013}
A.~K. Geim and I.~V. Grigorieva.
\newblock Van der waals heterostructures.
\newblock {\em Nature}, 499:419--425, 2013.

\bibitem{Novoselov2016}
K.~S. Novoselov, A.~Mishchenko, A.~Carvalho, and A.~H.~Castro Neto.
\newblock 2d materials and van der waals heterostructures.
\newblock {\em Science}, 353(6298):aac9439, 2016.

\bibitem{Feng2021}
Feng He, Yongjian Zhou, Zefang Ye, Sang-Hyeok Cho, Jihoon Jeong, Xianghai Meng, and Yaguo Wang.
\newblock Moiré patterns in 2d materials: A review.
\newblock {\em ACS Nano}, 15:5944--5958, 2021.

\bibitem{Wang2019}
J.~Wang, X.~Mu, L.~Wang, and M.~Sun.
\newblock Properties and applications of new superlattice: twisted bilayer graphene.
\newblock {\em Mater. Today Phys.}, 9:100099, 2019.

\bibitem{Andrei2020}
Eva~Y. Andrei and Allan~H. MacDonald.
\newblock Graphene bilayers with a twist.
\newblock {\em Nat. Mater.}, 19:1265--1275, 2020.

\bibitem{Yoo2019}
Hyobin Yoo, Rebecca Engelke, Stephen Carr, Shiang Fang, Kuan Zhang, Paul Cazeaux, Suk~Hyun Sung, Robert Hovden, Adam~W. Tsen, Takashi Taniguchi, Kenji Watanabe, Gyu-Chul Yi, Miyoung Kim, Mitchell Luskin, Ellad~B. Tadmor, Efthimios Kaxiras, and Philip Kim.
\newblock Atomic and electronic reconstruction at the van der waals interface in twisted bilayer graphene.
\newblock {\em Nature Materials}, 18:448--453, 2019.

\bibitem{Sunku2018}
S.~S. Sunku, G.~X. Ni, B.~Y. Jiang, H.~Yoo, A.~Sternbach, A.~S. McLeod, T.~Stauber, L.~Xiong, T.~Taniguchi, K.~Watanabe, P.~Kim, M.~M. Fogler, and D.~N. Basov.
\newblock Photonic crystals for nano-light in moiré graphene superlattices.
\newblock {\em Science}, 362:1153--1156, 2018.

\bibitem{Pei2011}
Pei Zhao, Qin Zhang, Debdeep Jena, and Siyuranga~O. Koswatta.
\newblock Influence of metal–graphene contact on the operation and scalability of graphene field-effect transistors.
\newblock {\em IEEE Transactions on Electron Devices}, 58(9):3170--3178, 2011.

\bibitem{Salvador2012}
Salvador Barraza-Lopez, Markus Kindermann, and M.~Y. Chou.
\newblock Charge transport through graphene junctions with wetting metal leads.
\newblock {\em Nano Letters}, 12(7):3424--3430, 2012.

\bibitem{Do2012}
V.~Nam~Do and H.~Anh~Le.
\newblock Transport characteristics of graphene-metal interfaces.
\newblock {\em Applied Physics Letters}, 101(16):161605, 2012.

\bibitem{Houssa2019}
Michel Houssa, Konstantina Iordanidou, Ashish Dabral, Augustin Lu, Geoffrey Pourtois, Valeri Afanasiev, and André Stesmans.
\newblock Contact resistance at mos2-based 2d metal/semiconductor lateral heterojunctions.
\newblock {\em ACS Applied Nano Materials}, 2(2):760--766, 2019.

\bibitem{Sancho1984}
M~P~Lopez Sancho, J~M~Lopez Sancho, and J~Rubio.
\newblock Quick iterative scheme for the calculation of transfer matrices: application to mo (100).
\newblock {\em J. Phys. F: Met. Phys.}, 14(5):1205, may 1984.

\bibitem{MacKinnon1985}
A.~MacKinnon.
\newblock The calculation of transport properties and density of states of disordered solids.
\newblock {\em Z. Phys. B}, 59:385--390, 1985.

\bibitem{Umerski1997}
A.~Umerski.
\newblock Closed-form solutions to surface green's functions.
\newblock {\em Phys. Rev. B}, 55:5266--5275, 1997.

\bibitem{Rivas2003}
Cristian Rivas and Roger Lake.
\newblock Non-equilibrium green function implementation of boundary conditions for full band simulations of substrate-nanowire structures.
\newblock {\em physica status solidi (b)}, 239:94--102, 2003.

\bibitem{Rocha2006}
A.~R. Rocha, V.~M. Garc\'{\i}a-Su\'arez, S.~Bailey, C.~Lambert, J.~Ferrer, and S.~Sanvito.
\newblock Spin and molecular electronics in atomically generated orbital landscapes.
\newblock {\em Phys. Rev. B}, 73:085414, 2006.

\bibitem{Sascha2017}
Sascha Brück, Mauro Calderara, Mohammad~Hossein Bani-Hashemian, Joost VandeVondele, and Mathieu Luisier.
\newblock Efficient algorithms for large-scale quantum transport calculations.
\newblock {\em The Journal of Chemical Physics}, 147:074116, 2017.

\bibitem{Nguyen2021}
V~Hung Nguyen, D~Paszko, M~Lamparski, B~Van Troeye, V~Meunier, and J-C Charlier.
\newblock Electronic localization in small-angle twisted bilayer graphene.
\newblock {\em 2D Materials}, 8(3):035046, 2021.

\bibitem{Gadelha2021}
Andreij~C. Gadelha and et~al.
\newblock Localization of lattice dynamics in low-angle twisted bilayer graphene.
\newblock {\em Nature}, 590:405--409, 2021.

\bibitem{Nguyen2022}
V~Hung Nguyen, Trinh~X Hoang, and J-C Charlier.
\newblock Electronic properties of twisted multilayer graphene.
\newblock {\em Journal of Physics: Materials}, 5(3):034003, 2022.

\bibitem{Trambly2010}
G.~Trambly~de Laissardière, D.~Mayou, and L.~Magaud.
\newblock Localization of dirac electrons in rotated graphene bilayers.
\newblock {\em Nano Letters}, 10(3):804--808, 2010.

\bibitem{Nguyen2015}
V~Hung Nguyen and P~Dollfus.
\newblock Strain-induced modulation of dirac cones and van hove singularities in a twisted graphene bilayer.
\newblock {\em 2D Materials}, 2:035005, 2015.

\bibitem{MacDonald2011}
Rafi Bistritzer and Allan~H. MacDonald.
\newblock Moiré bands in twisted double-layer graphene.
\newblock {\em Proceedings of the National Academy of Sciences}, 108(30):12233--12237, 2011.

\bibitem{Cao2018}
Yuan Cao, Valla Fatemi, Shiang Fang, Kenji Watanabe, Takashi Taniguchi, Efthimios axiras, and Pablo Jarillo-Herrero.
\newblock Unconventional superconductivity in magic-angle graphene superlattices.
\newblock {\em Nature}, 556:43--50, 2018.

\bibitem{Brun2020}
B~Brun, N~Moreau, S~Somanchi, V-H Nguyen, A~MreńcandKolasińska, K~Watanabe, T~Taniguchi, J-C Charlier, C~Stampfer, and B~Hackens.
\newblock Optimizing dirac fermions quasi-confinement by potential smoothness engineering.
\newblock {\em 2D Materials}, 7(2):025037, mar 2020.

\bibitem{Young2009}
Andrea~F. Young and Philip Kim.
\newblock Quantum interference and klein tunnelling in graphene heterojunctions.
\newblock {\em Nature Physics}, 5:222--226, 2009.

\bibitem{Rickhaus2013}
Peter Rickhaus, Romain Maurand, Ming-Hao Liu, Markus Weiss, Klaus Richter, and Christian Schönenberger.
\newblock Ballistic interferences in suspended graphene.
\newblock {\em Nature Communications}, 4:2342, 2013.

\bibitem{Datta2000}
Supriyo Datta.
\newblock Nanoscale device modeling: the green’s function method.
\newblock {\em Superlattices and Microstructures}, 28(4):253--278, 2000.

\bibitem{TB-SIM}
TB-SIM code: $\mathrm{www.mem}$-$\mathrm{lab.fr/en/Pages/L_{-}SIM/Softwares/TB_{-}Sim.aspx}$.

\bibitem{NANOTCAD}
NanoTCAD~ViDES: $\mathrm{http}$:$\mathrm{//vides.nanotcad.com}$.

\bibitem{Nemec2007}
Norbert Nemec and Gianaurelio Cuniberti.
\newblock Hofstadter butterflies of bilayer graphene.
\newblock {\em Phys. Rev. B}, 75:201404, 2007.

\bibitem{Hasegawa2013}
Yasumasa Hasegawa and Mahito Kohmoto.
\newblock Periodic landau gauge and quantum hall effect in twisted bilayer graphene.
\newblock {\em Phys. Rev. B}, 88:125426, 2013.

\bibitem{Moon2012}
Pilkyung Moon and Mikito Koshino.
\newblock Energy spectrum and quantum hall effect in twisted bilayer graphene.
\newblock {\em Phys. Rev. B}, 85:195458, 2012.

\bibitem{Moon2020}
J.~A. Crosse, Naoto Nakatsuji, Mikito Koshino, and Pilkyung Moon.
\newblock Hofstadter butterfly and the quantum hall effect in twisted double bilayer graphene.
\newblock {\em Phys. Rev. B}, 102:035421, 2020.

\bibitem{Wu2021}
QuanSheng Wu, Jianpeng Liu, Yifei Guan, and Oleg~V. Yazyev.
\newblock Landau levels as a probe for band topology in graphene moir\'e superlattices.
\newblock {\em Phys. Rev. Lett.}, 126:056401, 2021.

\bibitem{Yin2017}
Long-Jing Yin, Ke-Ke Bai, Wen-Xiao Wang, Si-Yu Li, Yu~Zhang, and Lin He.
\newblock Landau quantization of dirac fermions in graphene and its multilayers.
\newblock {\em Frontiers of Physics}, 12:127208, 2017.

\bibitem{Hejazi2019}
Kasra Hejazi, Chunxiao Liu, and Leon Balents.
\newblock Landau levels in twisted bilayer graphene and semiclassical orbits.
\newblock {\em Phys. Rev. B}, 100:035115, 2019.

\end{thebibliography}

\end{document}